\documentclass[fleqn,usenatbib]{mnras}

\usepackage{newtxtext,newtxmath}

\usepackage[T1]{fontenc}

\DeclareRobustCommand{\VAN}[3]{#2}
\let\VANthebibliography\thebibliography
\def\thebibliography{\DeclareRobustCommand{\VAN}[3]{##3}\VANthebibliography}

\usepackage{graphicx}	
\usepackage{amsmath}	
\usepackage{xcolor}     
\usepackage[normalem]{ulem} 

\title[Deep learning view of type 2 Seyferts]{A post-merger enhancement only in star-forming Type 2 Seyfert galaxies: the deep learning view}

\author[M. S. Avirett-Mackenzie et al.]{
  M. S. Avirett-Mackenzie,$^{1}$\thanks{E-mail: msam23@bath.ac.uk}
C. Villforth,$^{1}$
M. Huertas-Company,$^{2,3,4}$
S. Wuyts,$^{1}$
D. M. Alexander,$^{5}$
\newauthor\ 
S. Bonoli,$^{6,7}$
A. Lapi,$^{8}$
I. E. Lopez,$^{9,10}$
C. Ramos Almeida,$^{2,3}$
F. Shankar$^{11}$
\\
$^{1}$Department of Physics, University of Bath, Claverton Down, Bath BA2 7AY, UK\\
$^{2}$Instituto de Astrof\'{i}sica de Canarias, Calle V\'{i}a L\'{a}ctea, s/n, E-38205 La Laguna, Tenerife, Spain\\
$^{3}$Departamento de Astrof\'{i}sica, Universidad de La Laguna, E-38206 La Laguna, Tenerife, Spain\\
$^{4}$LERMA, Observatoire de Paris, CNRS, PSL, Universit\'{e} Paris Diderot F-75013, France\\
$^{5}$Centre for Extragalactic Astronomy, Department of Physics, Durham University, Durham DH1 3LE, UK\\
$^{6}$Donostia International Physics Center (DIPC), Manuel Lardizabal Ibilbidea, 4, E-20018 Donostia-San Sebasti\'{a}n, Spain\\
$^{7}$IKERBASQUE, Basque Foundation for Science, E-48013 Bilbao, Spain\\
$^{8}$SISSA, Via Bonomea 265, I-34136 Trieste, Italy\\
$^{9}$Dipartimento de Fisica e Astronomia `Augusto Righi', Universit\`{a} di Bologna, Via Gobetti 93/2, I-40129 Bologna, Italy\\
$^{10}$INAF -- Osservatorio di Astrofisica e Scienz dello Spazio di Bologna, Via Gobetti, 93/3, I-40129 Bologna, Italy\\
$^{11}$School of Physics and Astronomy, University of Southampton, Highfield, Southampton SO17 1BJ, UK
}

\date{Accepted XXX. Received YYY; in original form ZZZ}

\pubyear{2023}

\begin{document}
\label{firstpage}
\pagerange{\pageref{firstpage}--\pageref{lastpage}}
\maketitle

\begin{abstract}
  Supermassive black holes require a reservoir of cold gas at the centre of their host galaxy in order to accrete and shine as active galactic nuclei (AGN). Major mergers have the ability to drive gas rapidly inwards, but observations trying to link mergers with AGN have found mixed results due to the difficulty of consistently identifying galaxy mergers in surveys. This study applies deep learning to this problem, using convolutional neural networks trained to identify simulated post-merger galaxies from survey-realistic imaging. This provides a fast and repeatable alternative to human visual inspection. Using this tool, we examine a sample of $\sim$8500 Seyfert 2 galaxies ($L[\mathrm{\textsc{Oiii}}] \sim 10^{38.5 - 42}$ erg s$^{-1}$) at $z < 0.3$ in the Sloan Digital Sky Survey and find a merger fraction of $2.19_{-0.17}^{+0.21}\%$ compared with inactive control galaxies, in which we find a merger fraction of $2.96_{-0.20}^{+0.26}\%$, indicating an overall lack of mergers among AGN hosts compared with controls. However, matching the controls to the AGN hosts in stellar mass and star formation rate reveals that AGN hosts in the star-forming blue cloud exhibit a $\sim 2\times$ merger enhancement over controls, while those in the quiescent red sequence have significantly lower relative merger fractions, leading to the observed overall deficit due to the differing $M_{\ast} - {\rm SFR}$ distributions. We conclude that while mergers are not the dominant trigger of all low-luminosity, obscured AGN activity in the nearby Universe, they are more important to AGN fuelling in galaxies with higher cold gas mass fractions as traced through star formation. 
\end{abstract}




\section{\label{sec:introduction}Introduction}



It is well known 
that most galaxies host supermassive black holes (SMBHs) at their centres. Physical models~\citep[e.g.][]{silk+rees,king03} show that galaxies and their SMBHs interact constantly. When a galaxy has a reservoir of cold gas at its centre, the SMBH may accrete this gas rapidly, releasing a tremendous amount of energy as light and potentially even outshining the starlight of the galaxy as an active galactic nucleus (AGN). Some fraction of this energy is injected back into the gas surrounding the central region, either heating the gas or driving it out of the galaxy entirely. This in turn is expected to eventually quench star formation in the galaxy and starve out the AGN as the supply of cold gas required for both quickly disappears~\citep{di-matteo+08,hopkins+08}. Hence galaxies and their black holes are expected to grow together, as implied by the tight relationship between galaxy bulge mass and SMBH mass~\citep{kh13}.

Being located at the centre of a galaxy, an AGN must draw its supply of cold gas from within the galaxy. Many galaxies contain large reservoirs of such gas, but in order to reach the central region, gas orbiting at $r \sim 10$ kpc must lose $\gtrsim$99.9\% of its angular momentum~\citep{ah12}. Mechanisms for destabilising this gas may be environmental, such as gas-rich mergers~\citep{di-matteo+05, fontanot+15} or tidal interactions~\citep{mb08}, or internal, such as bar formation~\citep{sfb89, shankar+12} or wet compaction from violent disc instabilities at high redshifts~\citep{db14, zolotov+15, lapi+18, lapiner+23}. This work focuses on major mergers, which are known to drive large gas masses inwards particularly rapidly~\citep{cox+08, tcb10}.

In a merger between two gas-rich galaxies, the intense tidal forces and changing gravitational potential drive the gas to the centre of the system, where it may fragment and form stars or accrete on to the SMBH(s) to form an AGN, possibly initially shrouded in dust~\citep{sanders+88}. As the AGN accretion and starburst continue, the feedback from both will eventually shut both down as gas is either heated or blown out. Meanwhile, the merger will have redistributed stellar orbits into random orientations via violent relaxation~\citep{lb67}, and what is left at the end is a gas-poor, `red and dead' elliptical galaxy.

Whilst this scenario represents one possible formation pathway of an AGN, the question remains of how necessary mergers are for triggering AGN compared with other, less violent processes. Models indicate that some amount of merger triggering is necessary to match observed AGN demographics~\citep[e.g.][]{hkb14, db12}, but AGN have also been observed to exist in host galaxies with only a gas-rich disc~\citep[e.g.][]{cisternas+11, smethurst+19}, whose lack of a significant classical bulge component strongly suggests major-merger-free mass assembly histories. This implies that secular gas inflows from disc instabilities that build up over time can be sufficient to fuel SMBH growth in many cases.

A great deal of work has been done over the years seeking to pinpoint the relative importance of galaxy mergers and secular processes in triggering AGN activity, and results have been mixed. Many studies~\citep{cisternas+11, kocevski+12, villforth+14, sbh15, mechtley+16, marian+19, sharma+21} find no merger excess among AGN hosts compared with inactive control galaxies, suggesting that mergers have no additional contribution to AGN triggering. However, many other studies~\citep{ra+11, ra+12, ellison+19, marian+20, pierce+23} find overall merger excesses between their AGN hosts and controls. This discrepancy could be explained by the different types of AGN studied, as it has been suggested that mergers may only be the dominant trigger of high-luminosity AGN while secular processes are sufficient at lower luminosity~\citep{somerville+08, hh09, hkb14}, which has been supported by models~\citep{marulli+08, bonoli+09, menci+14, steinborn+18}, though observational evidence remains mixed~\citep{ah12, glikman+15, mechtley+16, villforth+17}. Further, the evolutionary model of AGN obscuration~\citep{sanders+88} predicts that obscured AGN occur more recently after the triggering event and thus may be more strongly associated with merger features, though again observations find mixed results~\citep{ulb08, satyapal+14, glikman+15, kocevski+15, donley+18, villforth+19}. Redshift may also play a role, as some models have suggested that mergers are more important for AGN triggering in the earlier Universe, with the dominant mechanism switching to secular processes at lower redshift~\citep{db12, menci+14}. Conversely, cosmological surface brightness dimming means that for a fixed surface brightness limit in a survey, faint merger features are less likely to be picked up at increasing redshift. Indeed,~\citet{pierce+23} found a positive correlation between imaging depth and observed merger excess, suggesting that inconsistent sensitivity in the images used to identify mergers may go a long way towards explaining the inconsistent results found over the years.

A further issue lies in the methods used to identify mergers. For the past century~\citep{hubble26}, many studies of galaxy morphology have relied on visual classification. The human eye is well-suited to identifying morphological features, including merger signatures, but this method is intrinsically subjective. While the basic Hubble types of spheroid vs disc are relatively well-defined, mergers are particularly tricky to define consistently. Comparing a few of the visual classification systems used in past studies gives a sense of the variety in the literature. \citet{cisternas+11} used three flags of increasing distortion level (none vs minor vs strong), which are chosen subjectively by the classifiers. \citet{ellison+19} used binary, non-exclusive flags marking disturbances and presence of a neighbour, though only the disturbance flag was considered as `interacting' in their analysis. \citet{ra+11} and further works from that group used the most complicated system with separate flags for different possible signatures (e.g. tidal tails, shells, etc). Interestingly, these three examples exhibit a trend of increasing observed merger fraction with increasing classification system complexity. Perhaps having a larger variety of features to consider may naturally lead the human brain to identify more occurrences. 

There is also the simple issue of visual inspection being time consuming, with the analysis of even a small dataset taking up many hours of an expert's precious research time. Crowd-sourcing projects such as Galaxy Zoo~\citep{gz-description} mitigate this issue (and also avoid some of the variation caused by subjectivity) by spreading the visual inspection work across thousands of citizen scientists and combining their votes into final classifications, but accuracy can suffer due to the lack of professional training of the volunteers~\citep[Galaxy Zoo classifiers were particularly hesitant to classify any galaxy as a merger; see][]{darg+10}. Further, as future surveys collect exponentially increasing volumes of data~\citep[e.g. LSST,][]{lsst}, even having an army of volunteers providing classifications is becoming increasingly impractical. 

Automatic morphological classification algorithms, on the other hand, are able to handle large volumes of data with ease, but traditionally have been seen as less accurate compared with human classifiers.\footnote{There is of course no ground truth to compare with for observations, but human classification is often seen as the gold standard, and most automatic classifiers have failed to agree with humans. It is important to note that the two methods of classification are looking for intrinsically different patterns, and galaxy morphology has historically been understood through the lens of human perception.}
This is due to the fact that until recently, most automatic classifiers have been based on measuring properties of the galaxy's shape and light distribution, such as concentration, asymmetry, clumpiness, Gini coefficient, and $M_{20}$. The galaxies are then divided in parameter space, either by taking cuts on each parameter~\citep[e.g.][]{abraham+96, cas, lotz+08} or using simple machine learning algorithms~\citep[e.g.][]{scarlata+07, hc+08,rose+23}. The values of these properties have been shown to vary with signal-to-noise~\citep{hc+14}, making their classifications inconsistent across surveys and failing at high redshifts~\citep{abruzzo+18}.

Over the last few years, deep learning, specifically with convolutional neural networks (CNNs) has seen much success in classifying general galaxy morphology~\citep[e.g.][]{hc+15, ds+18, cheng+20, spindler+21, hcl23} as well as identifying galaxy mergers~\citep[e.g.][]{bottrell+19b,pearson+19, wang+20, ciprijanovic+20, ciprijanovic+21,bickley+21, koppula+21}. As an automatic classification method, deep learning is easily repeatable and able to handle large data volumes far better than any number of human volunteers. CNNs have also been shown to outperform older automatic models based on previously measured features~\citep{cheng+20}, as they work by including image convolutions within the deep layers, where the weights trained form the convolutional kernels. Hence, during training they learn not only which features correspond to different labels but also what the features themselves are and where they are located in the images.

This work aims to resolve some of the ambiguity around the AGN-merger relationship through applying deep learning methods to study the merger status of a large sample of AGN host galaxies. We focus here on Type 2 AGN to avoid point source contamination, which hinders morphological analysis of the host~\citep[see][]{marian+19, villforth+19}. We also focus specifically on identifying post-mergers, which are thought to be the phase where black hole accretion rates peak~\citep{hopkins+08, blecha+18}, and which specifically require morphological (or kinematic) analysis (as opposed to pre-mergers, which can be identified as close pairs). We take the approach of supervised learning, where the neural network is trained to recognize mergers in a set of images of galaxies whose true merger status is known. This makes use of the IllustrisTNG cosmological simulation~\citep{tng-model-1, tng-model-2}, whose public data release includes merger trees, from which we assemble a catalogue of post-mergers and nonmergers, and visualisation tools, with which we create survey-realistic mock observations of galaxies in our catalogue. We then apply this classifier to a sample of Seyfert 2 host galaxies imaged by the Sloan Digital Sky Survey~\citep[SDSS;][]{sdss-dr7}, comparing their merger fractions to a mass- and redshift-matched control sample of inactive galaxies.

This paper is organized as follows. Section~\ref{subsec:data-sdss} describes our SDSS AGN hosts and inactive control galaxies, while Section~\ref{subsec:data-tng} introduces the simulated data we use to train the neural networks. Section~\ref{sec:methods} details the methods used to generate the training images and train the CNNs, and Section~\ref{sec:results} shows the results of applying the CNNs to classify the SDSS galaxies. Section~\ref{sec:discussion} interprets the results of the CNN classification, comparing to other studies and discussing caveats of our approach. Finally, Section~\ref{sec:conclusions} concludes. Consistent with IllustrisTNG, this paper assumes the Planck 2015 cosmology~\citep{planck15}.

\section{Data}
\label{sec:data}

\subsection{\label{subsec:data-sdss}SDSS AGN hosts}

\begin{figure}
  \centering
  \includegraphics[width=8cm]{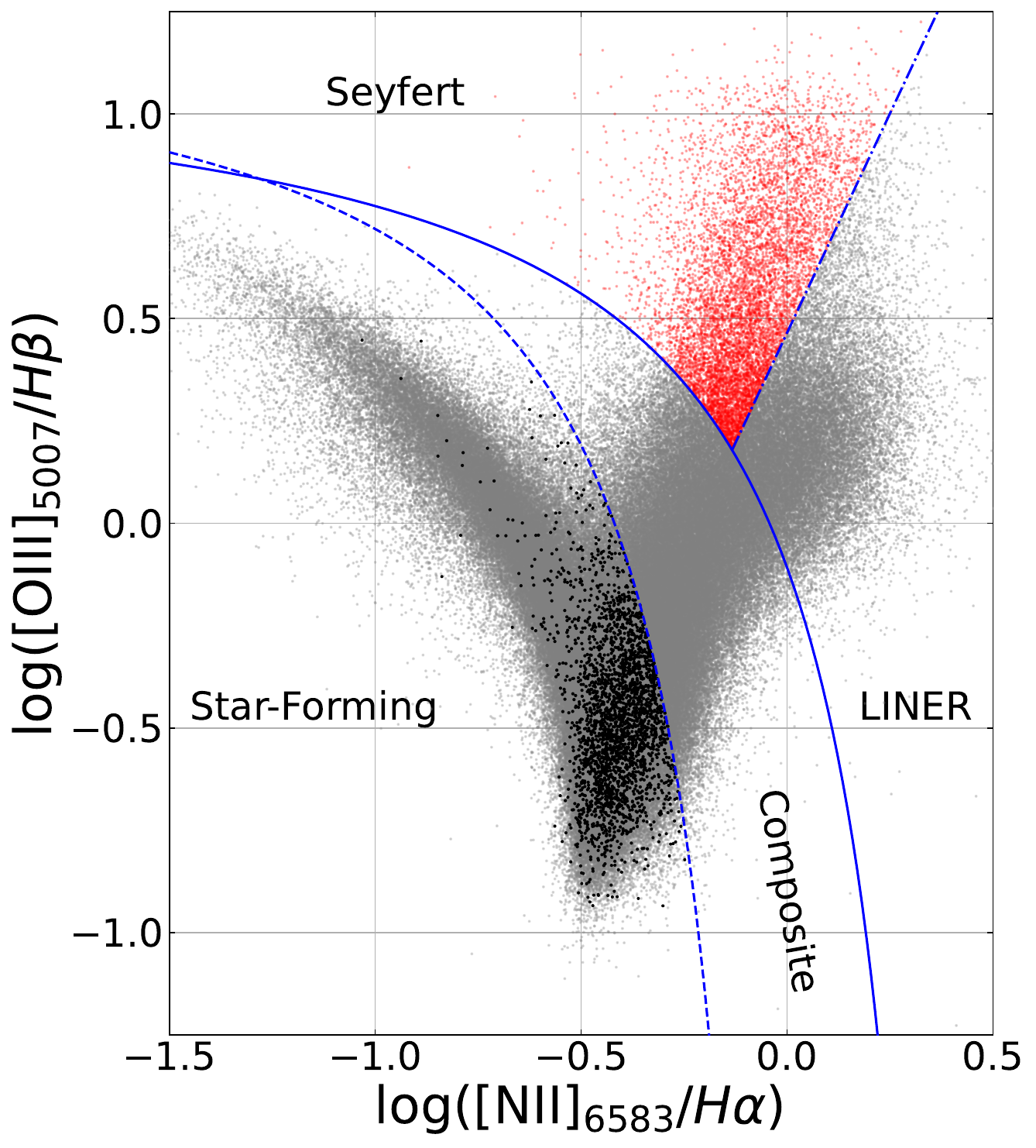}
  \caption{BPT diagnostics of our AGN sample (red) and the non-AGN galaxies in our sample with secure measurements of all four relevant emission lines (grey). Black points indicate the non-AGN selected as controls for this work (approximately half of the control sample, as the other half come from the set of galaxies without secure emission line detections and are not shown here). Galaxies are classified as star-forming below the~\citet{kauffmann+03-agn} cutoff (dashed line), as composite between this and the~\citet{kewley+01} cutoff (solid line), and as Seyfert or LINER above this and split by a straight line of angle $25^{\circ}$~\citep[dot-dashed line; see Figure 2 of][]{kauffmann+03-agn}.}
  \label{fig:bpt}
\end{figure}

This work considers merger fractions in Type 2 AGN only, so as to avoid point source contamination from unobscured AGN. We use the MPA-JHU catalogue~\citep{kauffmann+03-sf, brinchmann+04, salim+07} of SDSS DR7~\citep{sdss-dr7}, which contains emission line data and derived stellar masses and star formation rates of $\sim$800,000 galaxies, including AGN hosts. We consider all sources with well-constrained ($\sigma < 0.4$ dex) stellar masses greater than $10^9$ M$_{\odot}$ that are not classified as BROADLINE (Type 1) by the SDSS pipeline. To select our AGN sample, we use BPT emission line diagnostics~\citep{bpt, kewley+01, kauffmann+03-agn} on galaxies with a signal-to-noise ratio greater than 3 for their measurements of the H$\alpha$, H$\beta$, [\textsc{Oiii}] $\lambda 5007$, and [\textsc{Nii}] $\lambda 6583$ emission lines. The possible classes are star-forming, composite, Seyfert, and LINER as shown by Figure~\ref{fig:bpt}. We add to the Seyfert class galaxies with secure measurements of H$\alpha$, [\textsc{Oiii}], and [\textsc{Nii}] whose $3\sigma$ upper limit on H$\beta$ places them in the Seyfert region. Of the 14,979 galaxies classified as Seyferts in this way, 8492 have a sufficiently large angular size to consider them well-resolved, i.e. $R_{0.5 \mathrm{Petrosian}} > 1.5\times$FWHM of the field PSF in the $r$-band. These comprise our AGN sample.

\begin{figure}
  \centering
  \includegraphics[width=8cm]{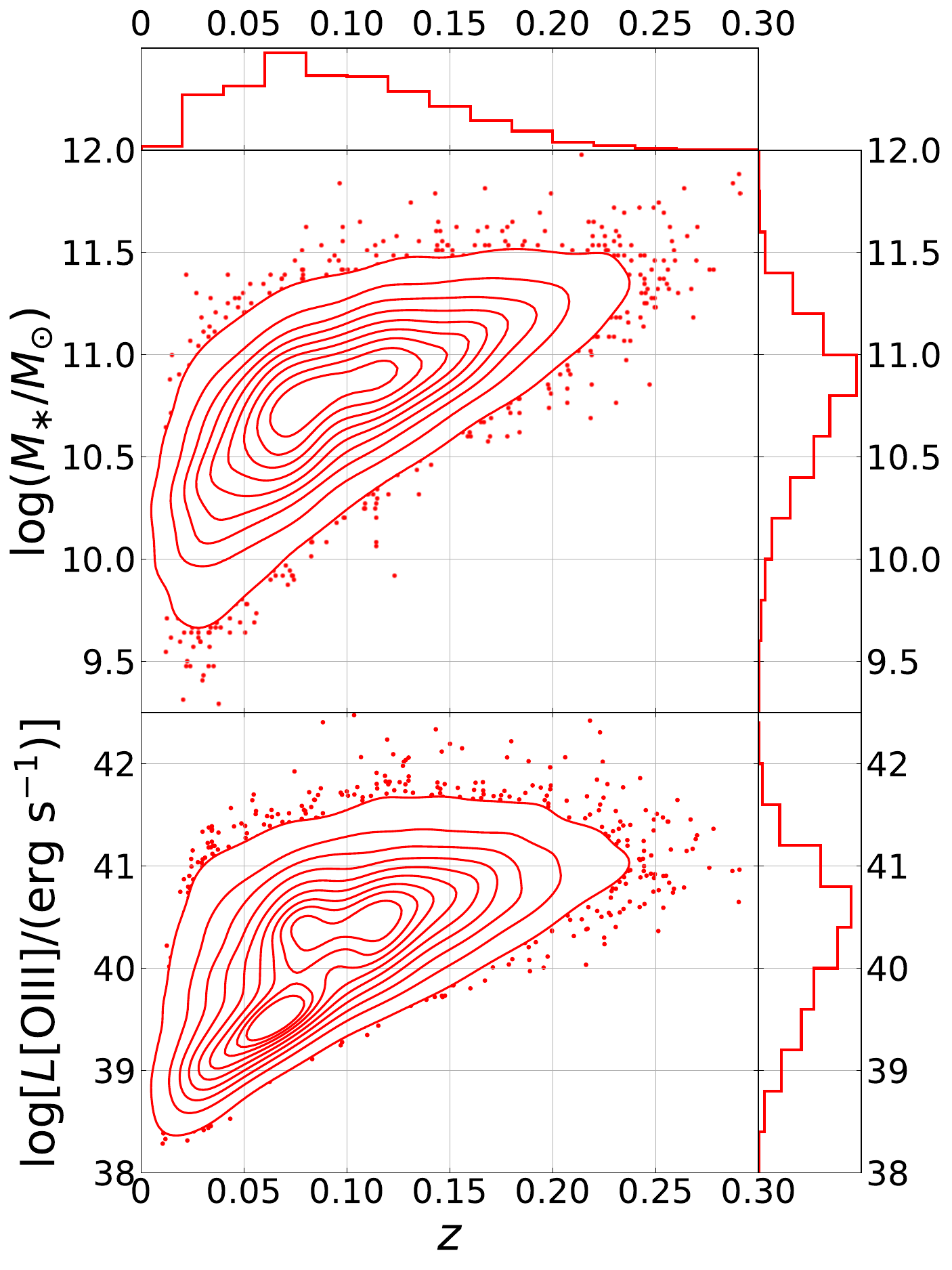}
  \caption{Distribution of our primary AGN sample (Seyfert 2 galaxies) in stellar mass (top) and [\textsc{Oiii}] luminosity versus redshift. The bin widths of the $M_{\ast}$ and $z$ histograms illustrate the bins used for control matching. Note that the control sample is matched in $M_{\ast}$ and $z$ and therefore the histograms for the control are identical and not shown here.}
  \label{fig:agn-mz}
\end{figure}

We select control galaxies from the BPT star-forming and low S/N (assumed majority quiescent) galaxies, excluding both composites and LINERs from the study. We match the control sample to the AGN sample by binning all well-resolved non-Seyfert galaxies (both those with non-Seyfert BPT classifications and those with low S/N emission lines) in mass and redshift (bin width 0.2 dex in $M_{\ast}$ and 0.02 in $z$, as shown in Figure~\ref{fig:agn-mz}). Then, for each bin in the AGN sample, we randomly select the same number of galaxies from the corresponding bin in the parent control sample. This ensures matching distributions, but individual AGN hosts do not have a single matched control galaxy. In total, the control sample selected this way is comprised of 32\% star-forming galaxies and 68\% low S/N (quiescent) galaxies.

\begin{figure}
  \centering
  \includegraphics[width=8cm]{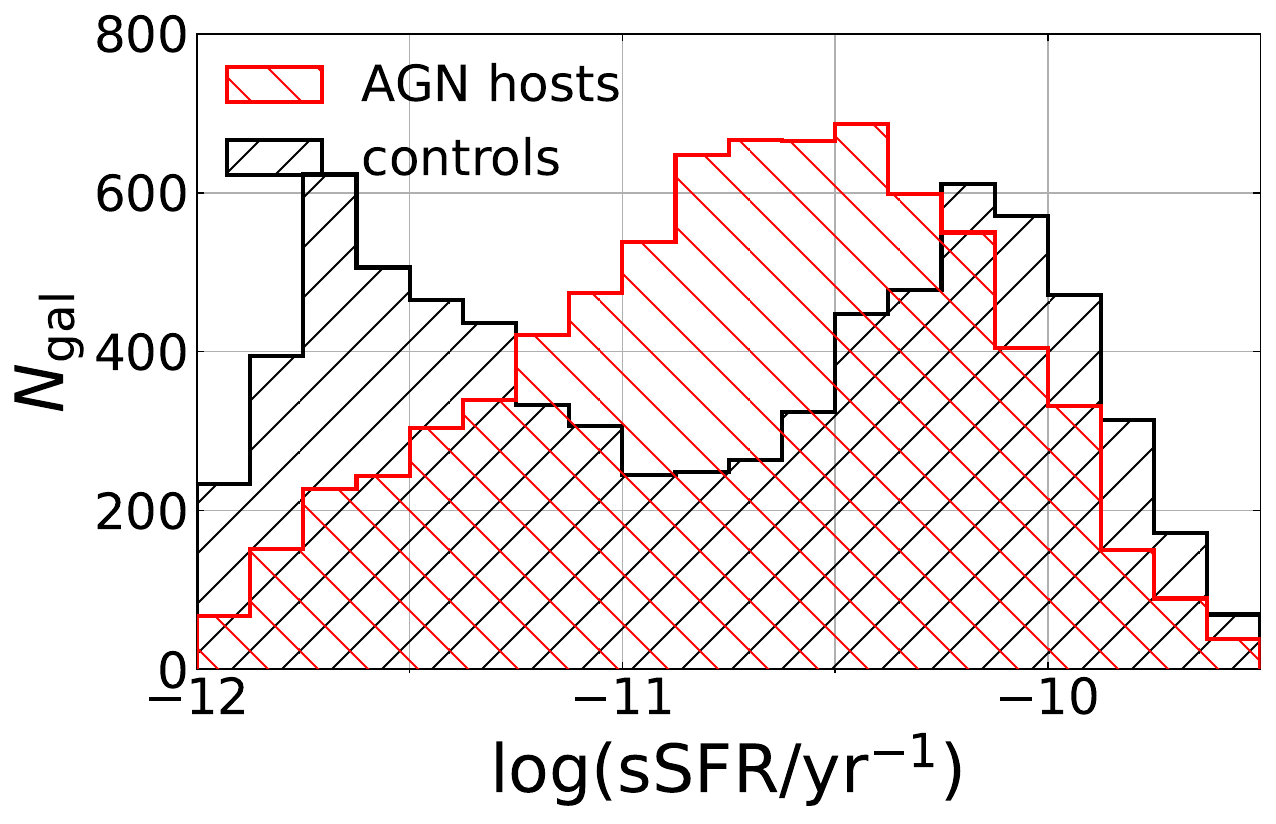}
  \caption{Distributions of specific star formation rates~\citep{brinchmann+04} of our AGN host sample (red) and their mass- and redshift-matched control galaxies (black). }
  \label{fig:sdss-ssfr}
\end{figure}

For the purpose of comparing observed merger fractions of subpopulations of the AGN and control samples divided by galaxy properties, we consider star formation rates and bulge-to-total fractions of our galaxies. \citet{brinchmann+04} derived star formation rates for the AGN in the MPA-JHU catalogue based on $D_N(4000)$, which avoids contamination from the AGN in the emission lines used to derive the SFR of inactive galaxies. Figure~\ref{fig:sdss-ssfr} shows the distributions of specific star formation rates (derived SFR/$M_{\ast}$) of our AGN hosts compared with the controls. We see that the control galaxies follow the typical bimodal distribution of star-forming and quiescent galaxies, while the AGN hosts follow a unimodal distribution peaking in the green valley. This is in line with observations that have shown AGN occupying galaxies at all stages of star formation but most often green valley and star-forming galaxies~\citep{cardamone+10, schawinski+10, aird+12, mullaney+15}.

We also examine trends with bulge-to-total fraction derived by~\citet{simard+11}, which may lend insight into the classification process of the CNN. 

\begin{figure*}
  \centering
  \includegraphics[width=14cm]{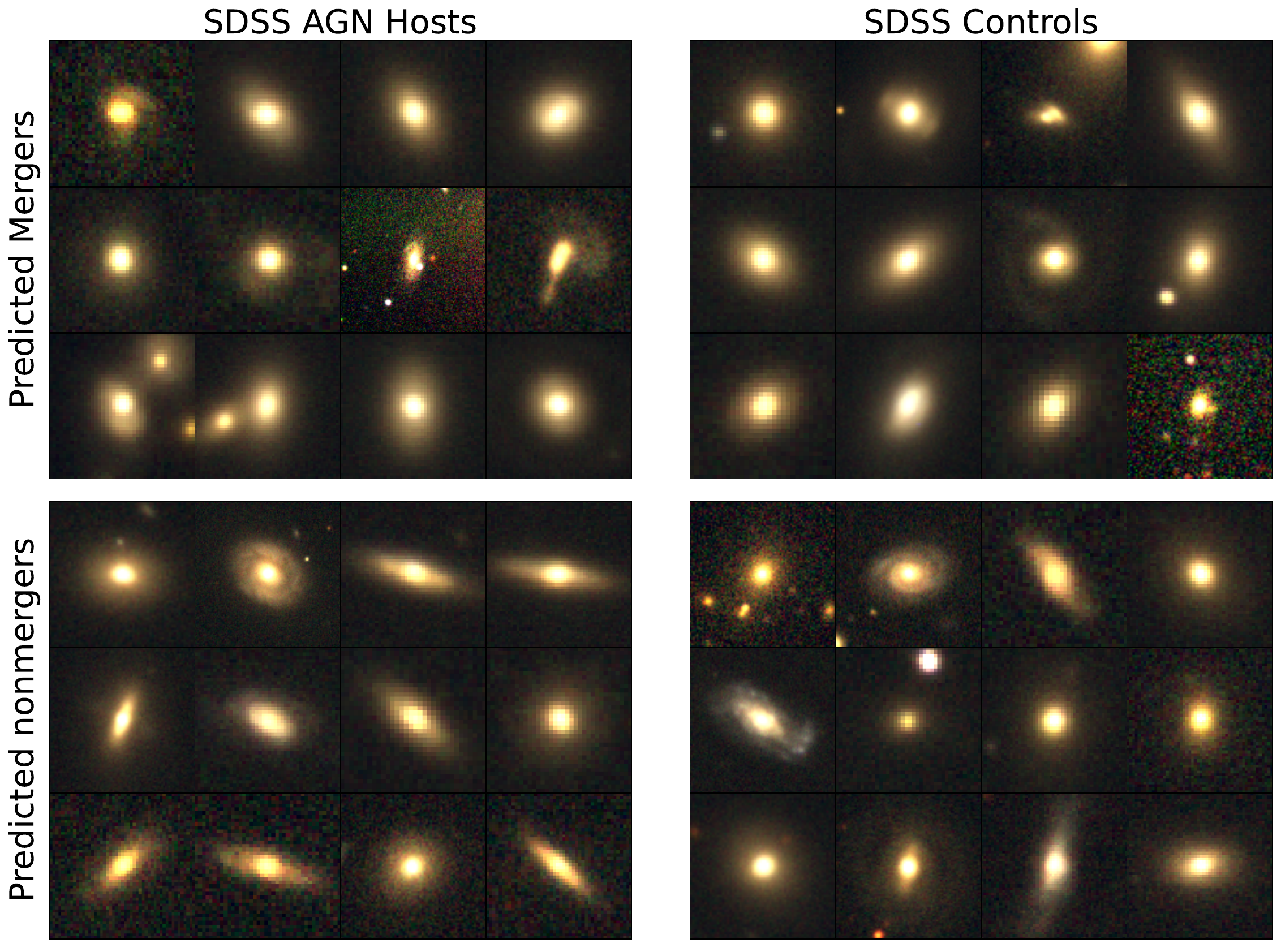}
  \caption{Example \emph{gri} cutouts of SDSS Seyfert 2s (left) and control galaxies (right), sorted into most confident mergers (top) and most confident nonmergers (bottom), as identified by our CNN ensemble described in Section~\ref{subsec:methods-ml}. Here, `most confident mergers' are galaxies classified as mergers by at least 84\% of the networks in the ensemble, while `most confident nonmergers' are classified as nonmergers by 100\% of the networks. Images are logarithmically scaled to better appreciate faint features, and the brightest 1\% of image pixels are saturated. Some of the predicted mergers have visually clear merger signatures or close companions, while others appear smooth and elliptical. Conversely, many of the predicted nonmergers are disc-dominated and some have asymmetries and features that could be interpreted visually as merger signatures or simply spiral arms.}
  \label{fig:sdss-ex}
\end{figure*}

For merger identification, we create a cutout image of each galaxy in the SDSS $gri$ bands. The cutouts are scaled to have a width of 8 times the Petrosian half light radius in the $r$-band, so that each galaxy occupies approximately the same area within its cutout (images are later rebinned to a uniform number of pixels for ML). Figure~\ref{fig:sdss-ex} shows example cutouts of AGN hosts and control galaxies. It is visually evident that galaxies span a variety of morphological types, and there is no obvious point source at the centres of the host galaxies.

\subsection{\label{subsec:data-tng}IllustrisTNG training data}

To build a deep learning-based classifier of galaxy morphology, we use a supervised learning approach with a training set drawn from the IllustrisTNG cosmological magnetohydrodynamical simulations, specifically the TNG100-1 simulation~\citep[hereafter referred to simply as TNG100]{tng-model-1, tng-model-2}. We use TNG because it simulates galaxy evolution in a cosmological context, including a variety of galaxies in different environments and thus giving a reasonably realistic approximation of the population of galaxies imaged in our SDSS sample. TNG has been shown to reproduce well most observed scaling relations of the galaxy population over cosmic time~\citep{tng-results-1, tng-results-2, tng-results-3, tng-results-4, tng-results-5}, 
as well as broadly representing the diversity seen in galaxy morphology~\citep{hc+19}. TNG100 has a box size of $\sim$100 Mpc and stellar particle resolution of $\sim$10$^{6}$ $M_{\ast}$, which places it in the middle of the TNG suite for both properties. This allows us to create a reasonably large set of reasonably realistic training images. 

TNG data are saved in snapshots capturing the simulation state at specific times, which are unevenly sampled but spaced by approximately 100--200 Myr. For each snapshot, haloes (representing galaxy groups or clusters) and subhaloes (representing galaxies) are determined by running a friends-of-friends algorithm and the \textsc{SubFind} algorithm~\citep{springel+05} respectively. Subhaloes are then traced across snapshots via the \textsc{SubLink} algorithm~\citep{vrg+15}, which produces merger trees, linked lists where each subhalo's entry links to its progenitor(s) in the preceding snapshot and descendant in the following snapshot. A galaxy having multiple progenitors indicates that a merger must have taken place since the previous snapshot. Hence we assemble a catalogue of post-mergers by looking backwards along the merger trees of galaxies matching our observational sample in mass and redshift, described in more detail in Section~\ref{subsec:methods-merger-trees}. 

This approach is limited by the coarse time sampling in the simulation, as we cannot image galaxies many times along the merger process~\citep[as done by][]{bottrell+19b, koppula+21} nor get more precise than a several hundred Myr upper bound on the time since coalescence of the identified post-mergers. However, making the reasonable assumption that the exact time of coalescence for each merger occurs at a random point between the last snapshot in which \textsc{SubFind} identifies two subhaloes and the first snapshot in which it finds only one, a sample of galaxies selected for having undergone a merger-tree-selected merger within a fixed time window can be expected to cover the full range of times since coalescence. The coarse time sampling of TNG100 is the cost of the large diversity of galaxies and cosmological context present in the simulation, which cannot be provided by higher-time-resolution simulations of individual galaxy pairs colliding.

\section{Methods}
\label{sec:methods}
We train an ensemble of neural networks to classify the merger state of galaxies from their cutout images, using supervised learning with a training set of simulated galaxy images derived from IllustrisTNG. Section~\ref{subsec:methods-merger-trees} describes the selection of the training set of post-merger and nonmerging galaxies from the simulation merger trees, while Section~\ref{subsec:methods-mock-observations} describes how we generated realistic mock observations of these galaxies. Finally, Section~\ref{subsec:methods-ml} describes the neural network architecture, image preprocessing, and training procedure. All machine learning procedures are performed using the \textsc{Keras} package~\citep{keras} of the \textsc{TensorFlow} API~\citep{tensorflow}.

\subsection{\label{subsec:methods-merger-trees}Training sample selection}

For our supervised learning model, we need a training set of IllustrisTNG galaxies that have lingering merger features and thus have undergone a sufficiently recent and major merger (our `post-merger' sample) and a matched control set that likely have no merger features and thus have not undergone any significant mergers in a long time (our `nonmerger' sample). To assemble this set, we begin by quantifying the merger state of each galaxy with $M_{\ast} > 10^9 M_{\odot}$ in the last 22 snapshots of the TNG100 simulation ($z \lesssim 0.3$). For each galaxy, we search its merger tree for the most recent merger above stellar mass ratio $\mu = 0.01$ (following \citealt{vrg+15}, we define $\mu$ as the ratio of the progenitor galaxy masses at the time prior to coalescence at which the smaller progenitor reaches its maximum stellar mass, thereby avoiding $\mu$ decreasing due to mass transfer). If this merger exists, we also check the 5 snapshots before coalescence for any higher-mass-ratio mergers so that e.g. a galaxy having undergone a $\mu = 0.01$ merger 200 Myr in the past and a $\mu = 1$ merger 400 Myr in the past will be labelled a a major rather than a minor merger.

Using this catalogue, we select our post-merger sample to consist of all galaxies in the aforementioned mass and redshift range with a major merger ($\mu \geq 0.25$) found within the last 500 Myr, which consists of 1954 galaxies. We match the nonmerger sample out of galaxies which have not undergone a merger with $\mu > 0.01$ within the last 2 Gyr (chosen with the aim of avoiding training images with intermediate-mass-ratio mergers, which may exhibit ambiguous features). The matching is performed by binning the post-mergers in mass (bin width 0.2 dex) and snapshot (bin width 2, corresponding to $\Delta z \sim 0.27$) and drawing an equal number of controls from each bin, similarly to the binning of the observed sample detailed in Section~\ref{subsec:data-sdss}.

\begin{figure}
  \centering
  \includegraphics[width=8cm]{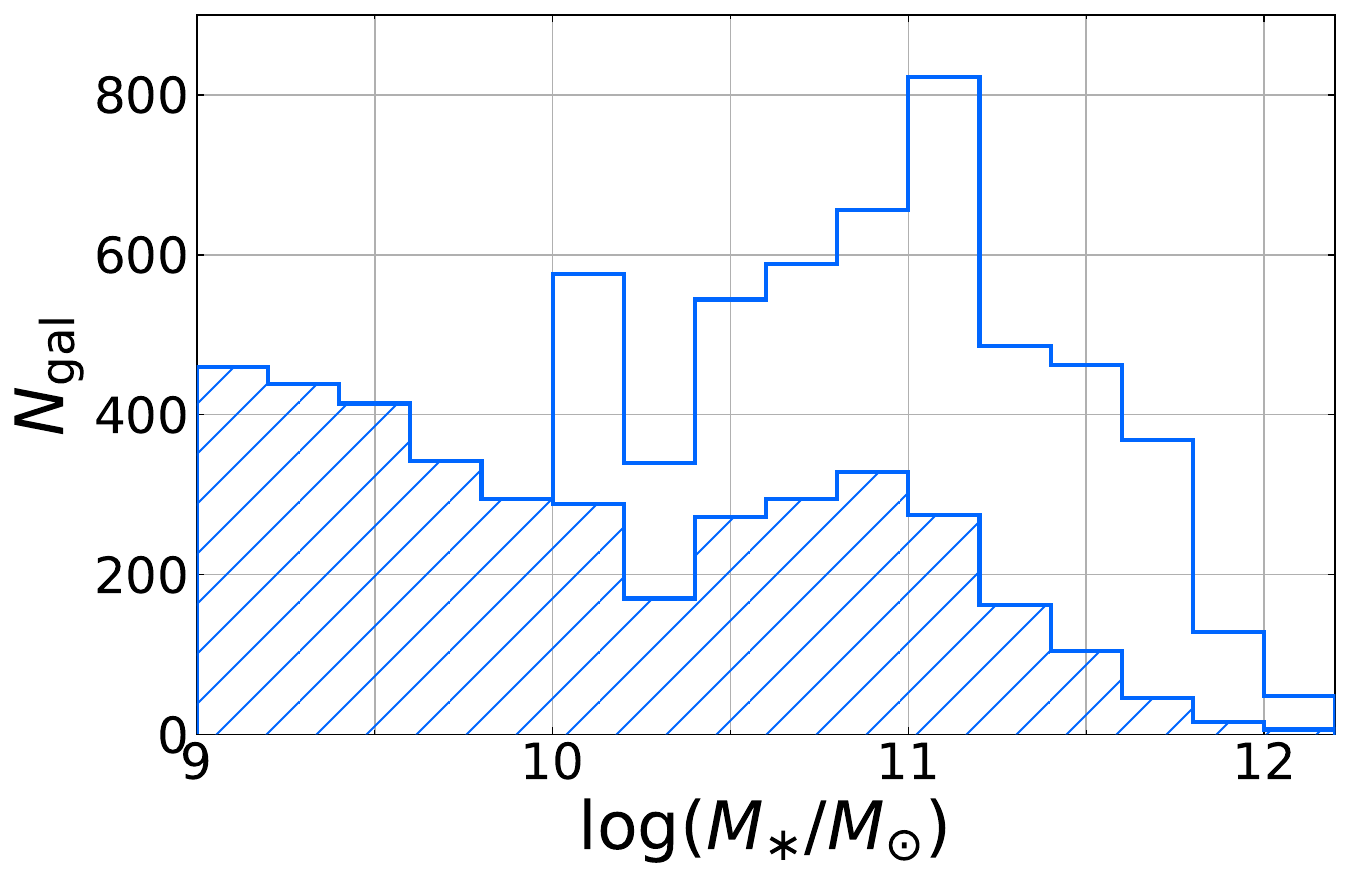}
  \caption{Mass distribution of the IllustrisTNG training galaxies. The hatched histogram shows the original distribution, while the open histogram shows the final sample we used after boosting the number of high-mass galaxy images with random image augmentations as described in Section~\ref{subsubsec:methods-image-prep}.}
  \label{fig:tng-masses}
\end{figure}

Figure~\ref{fig:tng-masses} shows the original stellar mass distribution of the training galaxies (hatched histogram). As this is the innate distribution of galaxies without observational selection effects, the fraction of low mass galaxies is much higher than in the SDSS sample, which creates issues with neural network prediction accuracy on galaxies with $M_{\ast} \gtrsim 10^{10.5} M_{\odot}$. 
To correct for this, we artificially increase the number of high-mass galaxy observations using image transformations in the d4 dihedral group (mirroring and rotations by $90^{\circ}$). 
Galaxies with $10 < \log(M_{\ast}/M_{\odot}) < 11$ have their number of observations doubled, while galaxies with $11 < \log(M_{\ast}/M_{\odot}) < 11.5$ have theirs tripled and galaxies with $\log(M_{\ast}/M_{\odot}) > 11.5$ have theirs increased by a factor of 8 (the maximum amount). This increases the effective number of training galaxies by a factor of $\sim$3, yielding the final boosted training sample whose mass distribution is shown in the open histogram of Figure~\ref{fig:tng-masses}.

\begin{figure*}
  \centering
  \includegraphics[width=16cm]{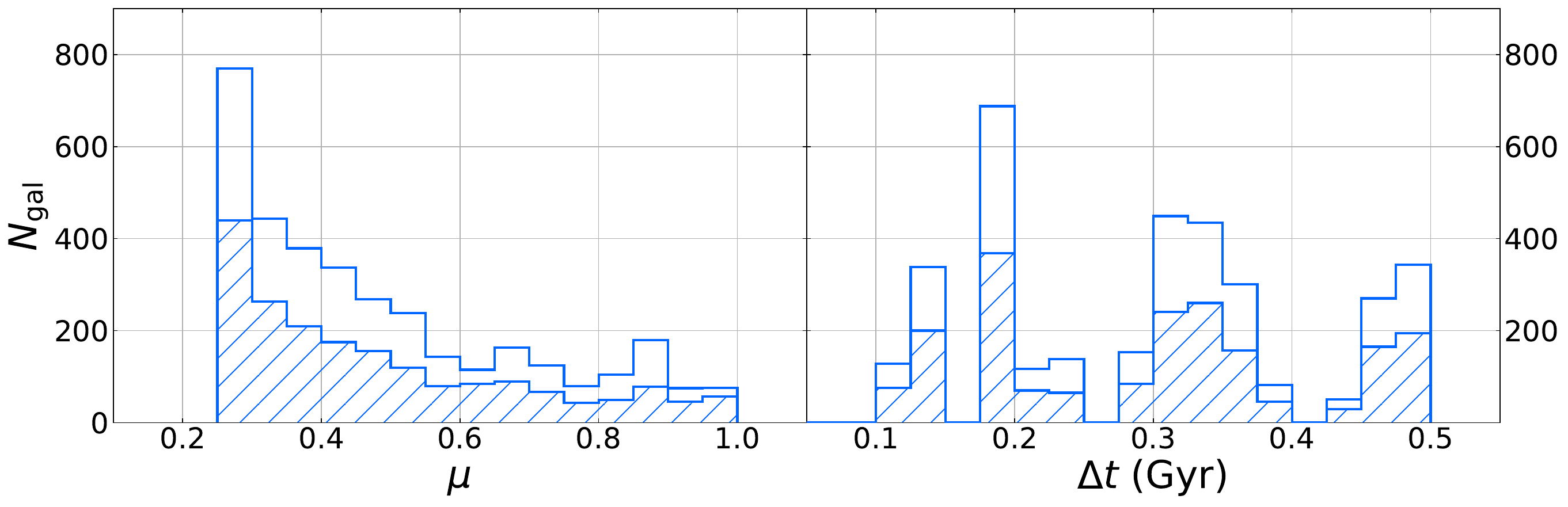}
  \caption{Mass ratio ($\mu$; left panel) and time since merger coalescence ($\Delta t$ right panel) distributions of the merger sample in the IllustrisTNG training set. Shaded and open regions represent the distributions before and after boosting the number of high-mass galaxies as described in Section~\ref{subsec:methods-merger-trees}.}
  \label{fig:mu-dt-dist}
\end{figure*}

Figure~\ref{fig:mu-dt-dist} shows the distribution of $\mu$ and $\Delta t$ in the merger training sample, where $\Delta t$ refers to the time difference between the imaged galaxy and the last snapshot where multiple progenitors are present. While the $\mu$ distribution falls off smoothly as expected, the $\Delta t$ distribution is highly clustered around certain time differences. This is due to the discrete time snapshots in the simulation data: the three peaks in the distribution indicate the time differences associated with one, two, and three snapshots since the merger (with spread due to the uneven time spacing of snapshots). We emphasize that $\Delta t$ represents the \emph{upper limit} on time since coalescence, as it could happen at any point between the snapshot with multiple progenitors and the snapshot with a single descendant, and the true distribution of time since coalescence is likely far more even (see Section~\ref{subsec:data-tng}). 

\subsection{\label{subsec:methods-mock-observations}Mock observations}

\subsubsection{\label{subsubsec:methods-idealised}Idealised image generation}

\begin{figure}
  \centering
  \includegraphics[width=7cm]{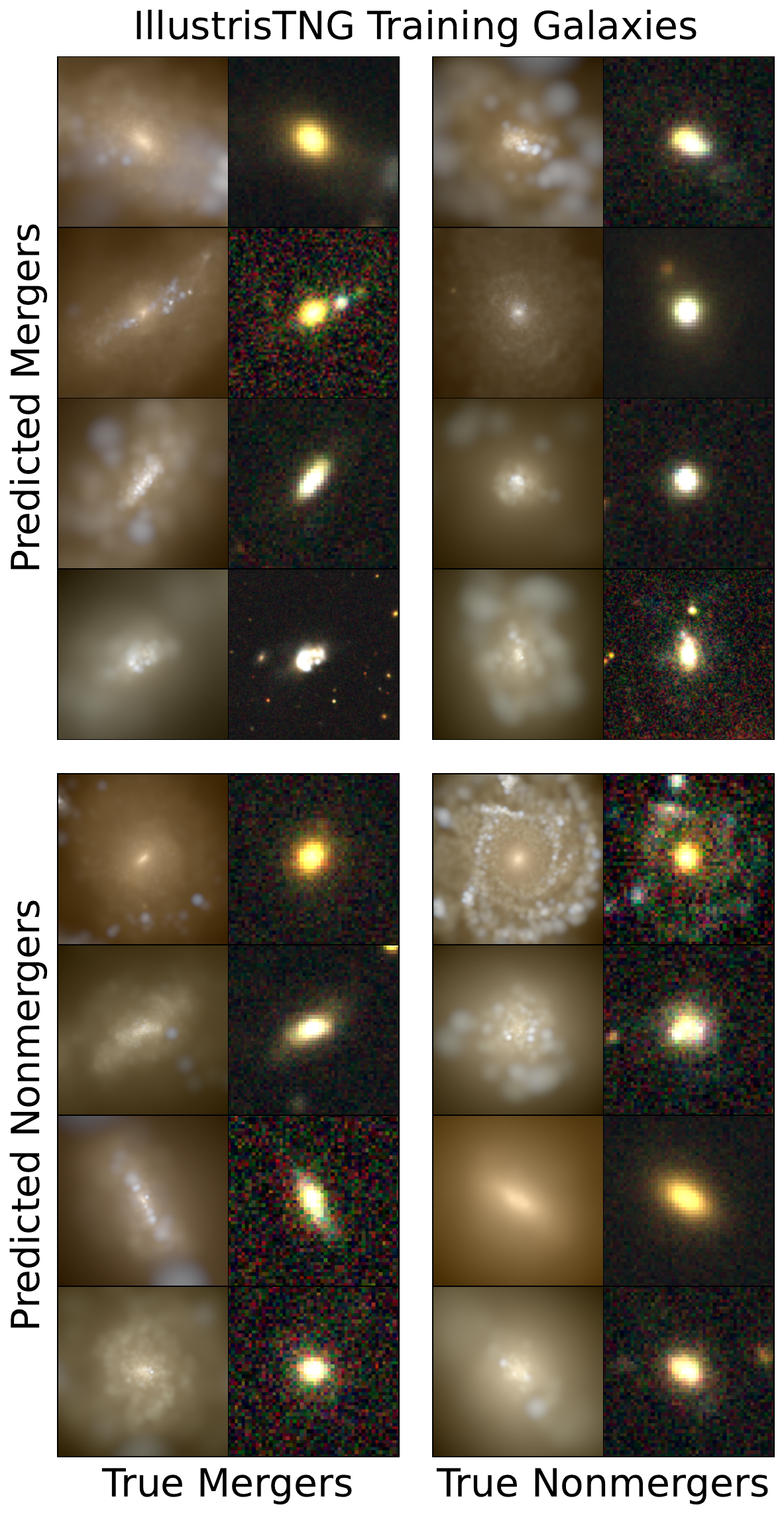}
  \caption{Example images from our IllustrisTNG training set sorted by true merger status (left panels: mergers; right panels: nonmergers) and CNN-predicted merger status (top panels: predicted mergers; bottom panels: predicted nonmergers). The left-hand images in each column show the idealised images described in Section~\ref{subsubsec:methods-idealised}, while the right-hand images show the same images after observational realism is applied (described in Section~\ref{subsubsec:methods-realism}). Images are logarithmically scaled to better appreciate faint features, and the realistic images have the brightest 1\% of their pixels saturated.}
  \label{fig:tng-ex}
\end{figure}

Images of the training sample are generated using the IllustrisTNG `Visualize Galaxies and Halos' online tool~\citep{tng-data}.\footnote{\url{https://www.tng-project.org/data/vis}} This tool projects all stellar particles associated with the galaxy and its parent halo within the field of view (including any other galaxies within the same halo), which we set to a width of 8 times the stellar half-mass radius $R_{0.5 M_{\ast}}$. It then uses the \textsc{fsps} stellar population synthesis model \citep{fsps1, fsps3} 
 and SDSS filter response functions~\citep{sdss-camera}
to generate an idealised mock observation of the galaxy. These visualisations do not include the effects of gas and dust in the ISM~\citep[which are not necessary to recover good merger identification, see][]{bottrell+19b}. Each galaxy is imaged from three orthogonal viewing angles, giving three independent observations per galaxy and hence tripling our training sample size to a total of $\sim$17000 images. The top panels of Figure~\ref{fig:tng-ex} show example idealised images of the TNG training mergers (left) and control nonmergers (right) in the SDSS \emph{gri} bands.

\subsubsection{\label{subsubsec:methods-realism}Observational realism}

In order to train a machine learning algorithm on mock observations of simulated galaxies and have confidence in its predictions on real observations, the noise and resolution properties of the training set must match the observed dataset as closely as possible. 
To achieve this, we apply the observational realism suite \textsc{RealSim} \citep{bottrell+17a, bottrell+17b, bottrell+19a, bottrell+19b}\footnote{\url{https://github.com/cbottrell/RealSim}} 
to the idealised images of our TNG training sample.
The aim of this algorithm is to transform idealised stellar-map-based images of galaxies into images indistinguishable from observations in scale, noise, and field properties. To accomplish this, the algorithm first rebins the image to the SDSS pixel scale, then convolves the idealised image with the PSF of a randomly-chosen SDSS field, then adds Poisson noise based on the SDSS exposure time of 53.9s, and finally inserts the image into an empty region of the same SDSS field to add the sky background. Figure~\ref{fig:tng-ex} shows the example images after realism is applied.\footnote{To test the effectiveness of \textsc{RealSim} in creating mock images indistinguishable from observations, two of the authors (MSAM and CV) carried out blind visual inspection on a subsample of random training images and SDSS cutouts. Neither could tell a difference in a majority of cases.}

As with the SDSS galaxies, an appreciable fraction of the training galaxies have radii too small to be well resolved by the SDSS telescope. This becomes an issue when realism is applied, since low resolution on images of mergers may lose identifying features and could bias the neural network to erroneously classify low-resolution images as mergers. In the interest of maintaining as large a training set as possible, rather than removing undersized galaxies, we place them at a lower redshift chosen to bring the image width up to the CNN input scale. While the IllustrisTNG galaxy visualisation tool will only create mock images in the rest frame or at each galaxy's original redshift, a redshift change of up to 0.3 does not significantly affect galaxy colour or the position of the 4000\AA\ break relative to the SDSS \emph{gri} bands. While in principle this could lower CNN performance on compact, higher-redshift galaxies, we note that similar galaxies were removed from the SDSS sample (see Section~\ref{subsec:data-sdss}), so our training galaxies continue to be representative of the science sample.

\subsection{\label{subsec:methods-ml}Neural network training and predictions}

\subsubsection{\label{subsubsec:methods-cnn}Neural network architecture}

We use a relatively simple convolutional neural network (CNN) architecture, similar in complexity to other CNNs currently used in the field~\citep[e.g.][]{bottrell+19b, ds+18, ciprijanovic+20}, but independently designed. A schematic of the CNN is shown in Table~\ref{table:simple-cnn-schematic}. The architecture consists of four blocks containing a convolutional layer with decreasing kernel sizes and increasing numbers of kernels, followed by batch normalisation and dropout layers. The third block includes a $2 \times 2$ maximum pooling layer, decreasing the image dimensions by half. The idea behind the varying convolutional kernel sizes and max pooling is to capture features at different scales within the image, while the batch normalisation and dropout layers are used to reduce overfitting. After these blocks follows a fully connected layer before the output layer, which is where classification is performed on the features extracted by the convolutional layers. The exact sizes of the convolutional kernels and fully-connected layer shown in Table~\ref{table:simple-cnn-schematic} were chosen via a hyperparameter search with \textsc{KerasTuner}~\citep{kerastuner}. The final output of the CNN is normalised to a merger probability, which we then convert to a binary prediction of merger/nonmerger by setting a decision threshold probability, which is explained in more detail in Section~\ref{subsec:results-cnn-training}.

\begin{table}
  \centering
  \begin{tabular}{lcc} \hline
    Layer                      & Output Shape  & \# Parameters \\ \hline \hline
    Input Layer                & (36, 36, 3)   & 0         \\ \hline
    Conv2D-1 $(6 \times 6)$    & (31, 31, 32)  & 3488      \\
    BatchNorm-1                & (31, 31, 32)  & 128       \\
    Dropout-1 $(0.1)$          & (31, 31, 32)  & 0         \\ \hline
    Conv2D-2 $(5 \times 5)$    & (27, 27, 64)  & 51,264    \\
    BatchNorm-2                & (27, 27, 64)  & 256       \\
    Dropout-2 $(0.1)$          & (27, 27, 64)  & 0         \\ \hline
    Conv2D-3 $(4 \times 4)$    & (24, 24, 128) & 131,200   \\
    BatchNorm-3                & (24, 24, 128) & 512       \\
    MaxPool-3 $(2 \times 2)$   & (12, 12, 128) & 0         \\
    Dropout-3 $(0.1)$          & (12, 12, 128) & 0         \\ \hline
    Conv2D-4 $(3 \times 3)$    & (10, 10, 128) & 147,584   \\
    BatchNorm-4                & (10, 10, 128) & 512       \\
    Dropout-4 $(0.1)$          & (10, 10, 128) & 0         \\ \hline
    Flatten                    & (12800)       & 0         \\
    Dense                      & (64)          & 819,264   \\ \hline
    Output Layer               & (2)           & 130       \\ \hline \hline
    Total Parameters           & $-$           & 1,154,338 \\ \hline
  \end{tabular}
  \caption{Schematic table showing the layers of the CNN used, as explained in Section~\ref{subsubsec:methods-cnn}.}
  \label{table:simple-cnn-schematic}
\end{table}

\subsubsection{\label{subsubsec:methods-image-prep}Image preprocessing}

The realistic TNG mock images and observed SDSS cutouts are preprocessed identically before being passed to the CNN for training and prediction. Images are first rebinned to a uniform $36 \times 36$ pixel scale to match the CNN input size. We estimate the sky level in each band by taking the median of pixels around the edge of the image (with a 3 px border thickness) and sky-subtract each band individually, then rescale the entire image cube logarithmically between 0 and 1 with a uniform background of 0.1 added. This process preserves intrinsic colour information in each galaxy.

The simulated images are randomly split 80/10/10 into training/validation/testing sets. The training set is the sample seen by the neural network during training, while the validation set is used to evaluate performance at each epoch and minimize overfitting. After training is complete, the testing set is used to judge the overall performance of the neural network at classifying galaxies into mergers and nonmergers, and all plots of performance metrics shown in this paper use the testing set. While the different viewing angles mean that the same galaxy may appear in both the training and testing sets, the mass boosting described in Section~\ref{subsec:methods-merger-trees} is applied only to the training and validation sets after splitting them. Thus, the same image in different orientations will not be repeated in multiple sets, and the performance metrics shown in Section~\ref{subsec:results-cnn-training} and Appendix~\ref{appendix:perf} accurately reflect the performance on the set of merging and nonmerging galaxies.

\subsubsection{\label{subsubsec:methods-training}Training procedure}

We generate merger fraction predictions using 100 CNNs in an ensemble, reducing the uncertainty arising from different random weight initialisations leading to different predictions between CNNs trained on the same dataset. We train each CNN for up to 500 epochs with a learning rate of $10^{-5}$, using the Adam optimiser~\citep{adam} and categorical crossentropy as the loss function (quantifying the difference between predicted merger probabilities and labels of 0 or 1). To minimise overfitting, training images undergo random reflections, rotations up to $15^{\circ}$, and shifts up to a few pixels at each epoch. These augmentations performed at training time are distinct from those performed to artifically increase the number of images of higher-mass galaxies: not only are the augmentation parameters different, but the aim here is to avoid CNNs learning a dependence on galaxy orientation, and training-time augmentations change randomly at each epoch. Training is also cut off if the loss function of the validation set fails to improve after 50 epochs, which typically indicates overfitting beginning to happen. The averaged training histories are shown in Figure~\ref{fig:histories}.

\section{Results}
\label{sec:results}

The trained ensemble of CNNs form a classifier with the ability to identify merger candidates and hence predict merger fractions within SDSS galaxy populations, estimating the uncertainty of its predictions arising from variations between different neural network instances. Section~\ref{subsec:results-cnn-training} describes the overall performance of this classifier on the testing subset of our IllustrisTNG mock images, where the merger state of each galaxy is known and thus prediction accuracy can be assessed. Section~\ref{subsec:results-merger-fraction} presents the results of applying this classifier to our sample of Type 2 AGN hosts and their matched controls, comparing merger fractions of the overall samples as well as examining potential trends with different galaxy and AGN properties. Section~\ref{subsec:results-validation} discusses caveats of our method and steps taken to validate the CNN predictions.

\subsection{\label{subsec:results-cnn-training}Neural network performance}

\begin{figure}
  \centering
  \includegraphics[width=8cm]{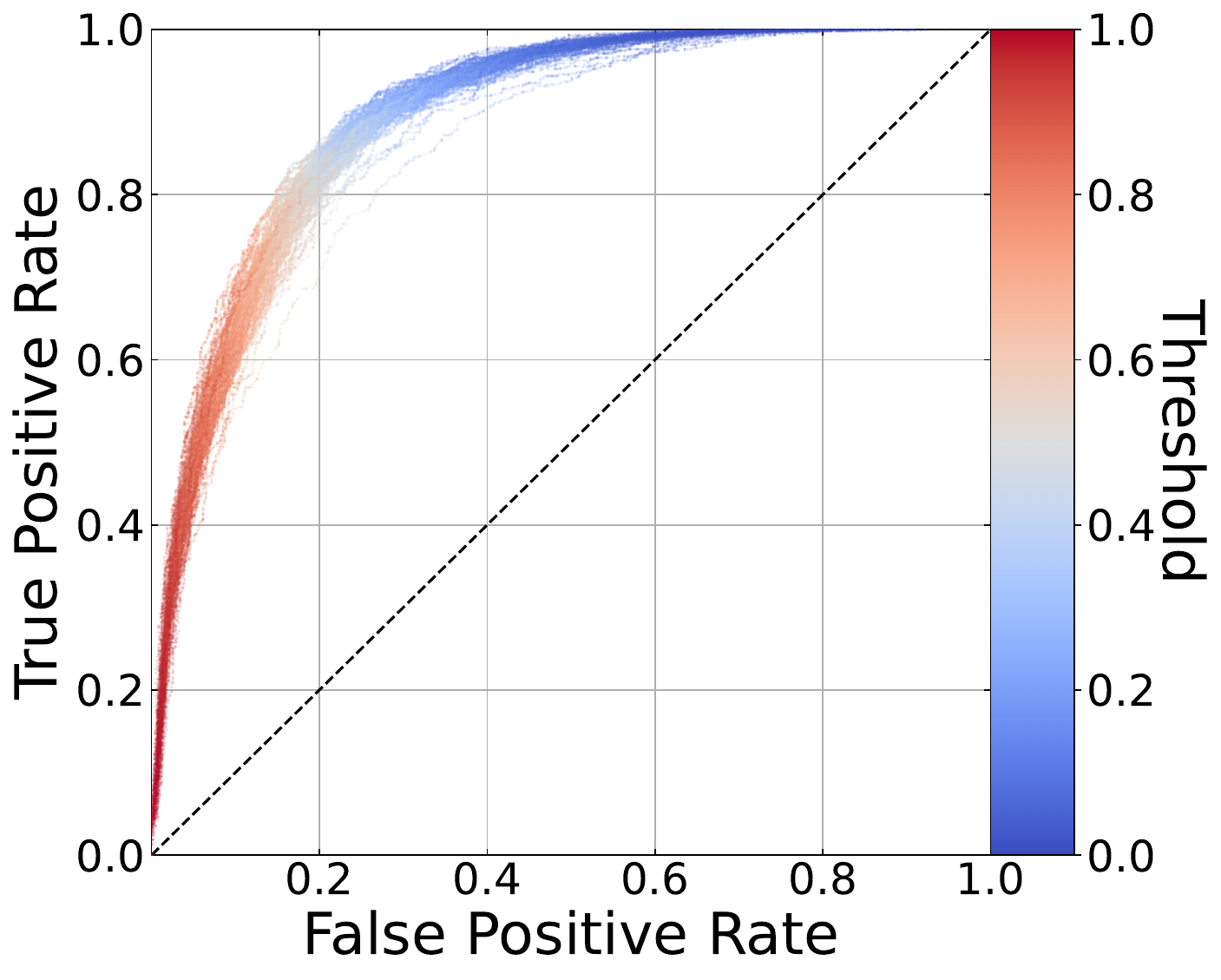}
  \caption{Receiver Operating Characteristic (ROC) curves of CNN performance on the testing set of the TNG galaxies, composed of true positive rate and false positive rate plotted at varying threshold probabilities between predicting a merger vs nonmerger (colour scale). The dashed black line represents a completely random classifier, while a ROC curve reaching the top left corner would represent a perfect classifier.}
  \label{fig:roc}
\end{figure}

Each CNN outputs a merger probability $P_{\mathrm{mg}}$ for each image. We may measure the overall effectiveness of each CNN by its receiver operating characteristic (ROC) curve, which is calculated by computing true positive rate (fraction of true mergers correctly identified; TP/(TP+FN)) and false positive rate (fraction of nonmergers falsely identified as mergers; FP/(FP+TN)) and varying the threshold probability below which a galaxy is classified as a nonmerger and above which it is classified as a merger. Figure~\ref{fig:roc} shows the ROC curves for each of our 100 CNNs on the testing set of the labelled TNG mock images. A completely random classifier would have a ROC curve following the straight dashed line in Figure~\ref{fig:roc}, while a perfect classifier would reach the top left corner, with a true positive rate of 100\% and a false positive rate of 0\%. Hence, we can use the area under the curve (AUC) as a metric of performance: our CNNs achieve a median AUC of $0.893_{-0.008}^{+0.005}$.

To convert the output probabilities to a predicted merger fraction, a fixed $P_{\mathrm{mg}}$ threshold must be set. Figure~\ref{fig:roc} shows that most of the CNNs achieve the shortest distance to 100\% true positives and 0\% false positives with a classification threshold close to 0.5, so taking this into consideration (along with the observation that the vast majority of the CNN $P_{\mathrm{mg}}$ values are well separated, i.e. close to either 0 or 1; see Figure~\ref{fig:p-mg-ex}), we set a uniform classification threshold of 0.5 for all CNNs. Repeating the measurements presented in Section~\ref{subsec:results-merger-fraction} with uniform thresholds between 0.4 and 0.6 showed no qualitative change to the scientific results, only a change in overall normalisation. Hence, given a set of galaxies (e.g. the AGN sample, or galaxies separated into bins as in Figures~\ref{fig:f-mg-1d} and~\ref{fig:enh-mstar-sfr}), the merger fraction is calculated as 
\begin{equation}
  f_{\mathrm{mg}} = \frac{\#(P_{\mathrm{mg}} > 0.5)}{\# \mbox{ galaxies in sample}}.
\end{equation}
This is calculated separately for each CNN. Merger fractions quoted in text and shown in figures are medians unless otherwise stated. Error bars are calculated by combining 1$\sigma$ binomial confidence intervals calculated using the method of~\citet{cameron11} with classification variations between different CNNs, represented by 16th and 84th percentiles of the 100 CNN predictions. Errors are generally dominated by the binomial term.

With the classification threshold fixed, our CNNs perform generally well, reaching a median accuracy of $80.8_{-0.8}^{+1.3}$\%. This is comparable to performances achieved in other recent works using CNNs to identify simulated mergers in realistic images~\citep{bottrell+19b, wang+20, ciprijanovic+21}. Precision (purity, fraction of true mergers out of galaxies identified as such; TP/(TP+FP)) and recall (completeness, fraction of correctly-classified true mergers; TP/(TP+FN)) fall into similar ranges of $ 80.9_{-1.2}^{+1.2}$\% and $81.1_{-2.4}^{+2.6}$\% respectively. Given that our sample here has been balanced to have an equal number of post-mergers and nonmergers, while in nature mergers are rare events, it is important to keep in mind that the imperfect accuracy and purity indicate that merger samples derived using these classifiers will contain a large number of false positives.

For the CNNs to reliably predict merger fractions for real galaxies, we must be sure they are basing their predictions on morphology and not becoming biased by galaxy properties that are unrelated or loosely correlated to merger state. To verify this, we bin the TNG testing set galaxies in stellar mass, redshift, specific star formation rate, and half mass radius and compare both prediction accuracy and predicted vs intrinsic merger fraction for each property.
We find no correlations between false positive rate and any of these observables, indicating no regions with systematically higher contamination rates. We do, however, observe that post-mergers with higher stellar masses and larger physical radii (both relatively rare) and lower specific star formation rates have a higher chance of being a false negative (see Figures~\ref{fig:fmg-tng} and~\ref{fig:fpr-tpr}).

\begin{figure*}
  \centering
  \includegraphics[width=17cm]{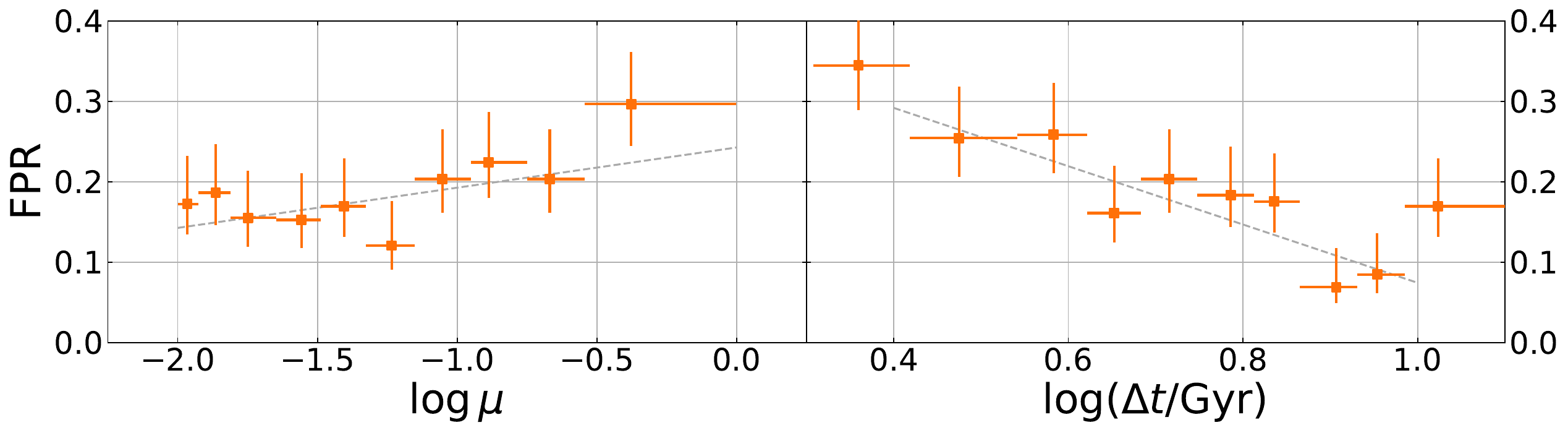}
  \caption{False positive rate (FPR) of our CNN ensemble as a function of mass ratio $\mu$ (left) and time since merger $Delta t$ (right). The TNG testing set is split into equally-filled bins to create these plots: horizontal error bars represent the bin widths. Grey dashed lines show linear best fits.}
  \label{fig:fpr-mu-dt}
\end{figure*}

To understand how merger state affects the CNN predictions, we consider true and false positive rate (TPR and FPR) as a function of mass ratio $\mu$ and time since merger $\Delta t$. A Pearson correlation test indicates that TPR shows no significant trends with either $\mu$ or $\Delta t$ 
($p = 0.17$ and 0.19 respectively): evidently, the CNNs are equally good at identifying mergers over the $\mu$ and $\Delta t$ ranges adopted. However, FPR shows a positive correlation ($r=0.61;\ p=0.0046$) with the mass ratio of the most recent merger with $\mu > 0.01$ and a negative one ($r=-0.85;\ p=0.0016$) with $\Delta t$ (see Figure~\ref{fig:fpr-mu-dt}). This shows that if a galaxy has undergone a major merger at any point, the CNNs are more likely to classify it as a post-merger. Further, the more recently a galaxy has undergone a merger of any mass ratio, the more likely it will be classified as a post-merger, a trend persisting even over timescales of several Gyr. This suggests that either merger-specific features linger in TNG for up to several Gyr (potentially depending on mass ratio), or the CNNs are basing classifications on features that linger for longer than asymmetry and tidal tails, such as bulge dominance.

\subsection{\label{subsec:results-merger-fraction}Merger fractions of Type 2 Seyferts and inactive galaxies}

With an unbiased classifier that identifies mergers with $\gtrsim$80\% accuracy, we begin our analysis of the merger-AGN relationship in the observed SDSS sample by comparing overall predicted merger fraction of the AGN hosts with the controls. We in fact find a decrease of median merger fraction in AGN hosts relative to controls, with $f_{\mathrm{mg, AGN}} = 2.19_{-0.17}^{+0.21}$\% and $f_{\mathrm{mg, control}} = 2.96_{-0.20}^{+0.26}$\%.

To quantify the significance of this result, we calculate the probability of an enhancement (${\rm enh} = f_{\rm mg,AGN} / f_{\rm mg,control}$) greater than 1 as follows. For each CNN in the ensemble, we randomly sample beta distributions based on the merger fraction measured by that CNN, one for the AGN hosts and one for the controls. Each pair of merger fractions drawn is divided to generate a distribution of enhancements, which we sum over the CNNs in the ensemble to create a combined pdf. We integrate the pdf to obtain $p({\rm enh} > 1)$. All enhancement significances quoted in this paper are calculated using 10000 draws for each CNN.

\begin{figure}
  \centering
  \includegraphics[width=8cm]{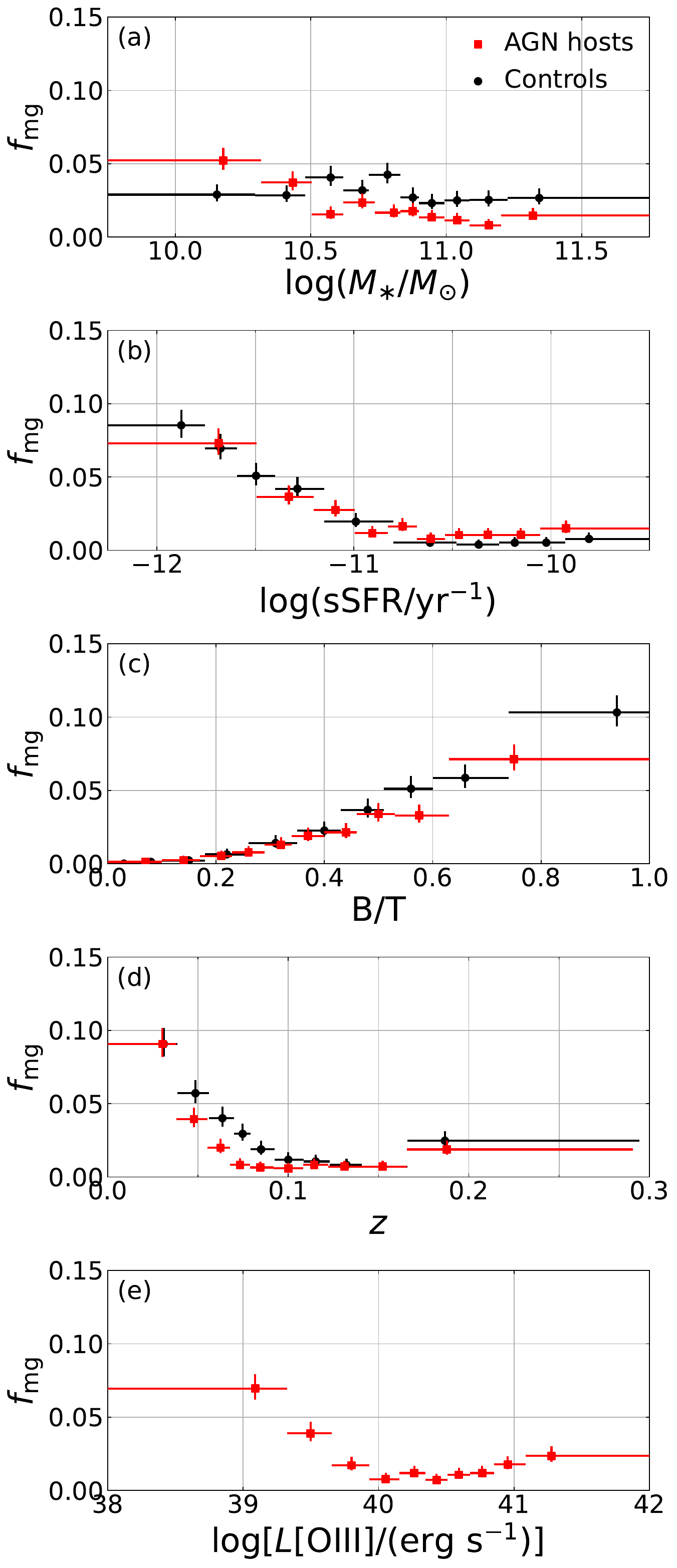}
  \caption{Merger fractions of SDSS AGN hosts (red squares) and control galaxies (black circles) binned in (a) stellar mass, (b) specific star formation rate, (c) bulge-to-total ratio~\citep{simard+11}, (d) redshift, and (e) [\textsc{Oiii}] luminosity (without controls as their [\textsc{Oiii}] luminosities are dominated by star formation rather than nuclear activity). Bins are chosen such that the total number of galaxies per sample is equal in each bin. The points represent the median value of each galaxy/AGN property and median merger fraction calculated for each bin. Horizontal error bars represent bin width, while vertical error bars represent $1\sigma$ confidence intervals on the merger fractions (as calculated from binomial errors and CNN variance; see Section~\ref{subsec:results-cnn-training}). While the controls are matched in $\log M_{\ast}$ and $z$ and thus these bins have equal edges, they are not matched in the other properties shown, notably sSFR (see Figure~\ref{fig:sdss-ssfr}). Note that the most extreme bins in both panels represent the tails of each property's distribution and hence are potentially very noisy.}
  \label{fig:f-mg-1d}
\end{figure}

   Using this method, we calculate that $p({\rm enh} > 1) = 0.0105$, indicating that our observation of a relative lack of mergers among AGN hosts overall is significant to just shy of the $3\sigma$ level. Hence, our first observation is that on a global scale, we see no evidence for mergers being the dominant trigger of all activity in obscured low-luminosity AGN in the nearby Universe. However, the picture becomes more nuanced when we separate our AGN hosts and controls by galaxy properties and compare observed merger fractions at fixed values of these properties. Figure~\ref{fig:f-mg-1d} shows predicted merger fractions of the two populations binned by properties of the stellar populations, bulge-to-total ratio, redshift, and AGN luminosity. For the comparison, we divide both AGN hosts and control galaxies into bins containing equal numbers of objects. 
The rest of this section is devoted to discussing the trends found.

\subsubsection{\label{subsubsec:f-mg-m-sfr}Merger fraction and stellar populations}

Panels (a) and (b) of Figure~\ref{fig:f-mg-1d} show predicted merger fractions of the AGN hosts compared with controls as a function of (a) stellar mass and (b) specific star formation rate. Comparing the binned data with a Pearson correlation test, the controls show no significant evolution of merger fraction with stellar mass, while the AGN hosts exhibit a correlation coefficient of $r = -0.84$ with high significance ($p = 1.6 \times 10^{-3}$). These combined trends result in a potential enhancement in the lowest-mass bin and relative lack of mergers among AGN hosts in the highest-mass bins.

Both populations show significant negative trends of $f_{\mathrm{mg}}$ with sSFR, though the trend for the AGN hosts is both flatter and of lower significance ($r = -0.81;\ p = 0.005$) compared with that of the controls ($r = -0.92;\ p = 1.6 \times 10^{-4}$). Comparing the two trends in Figure~\ref{fig:f-mg-1d} panel (b), this results in a merger enhancement among AGN hosts among galaxies with $\log(\mathrm{sSFR/yr}^{-1}) \gtrsim -10.5$, although the merger fraction among all galaxies in this region is very low.

\begin{figure}
  \centering
  \includegraphics[width=8cm]{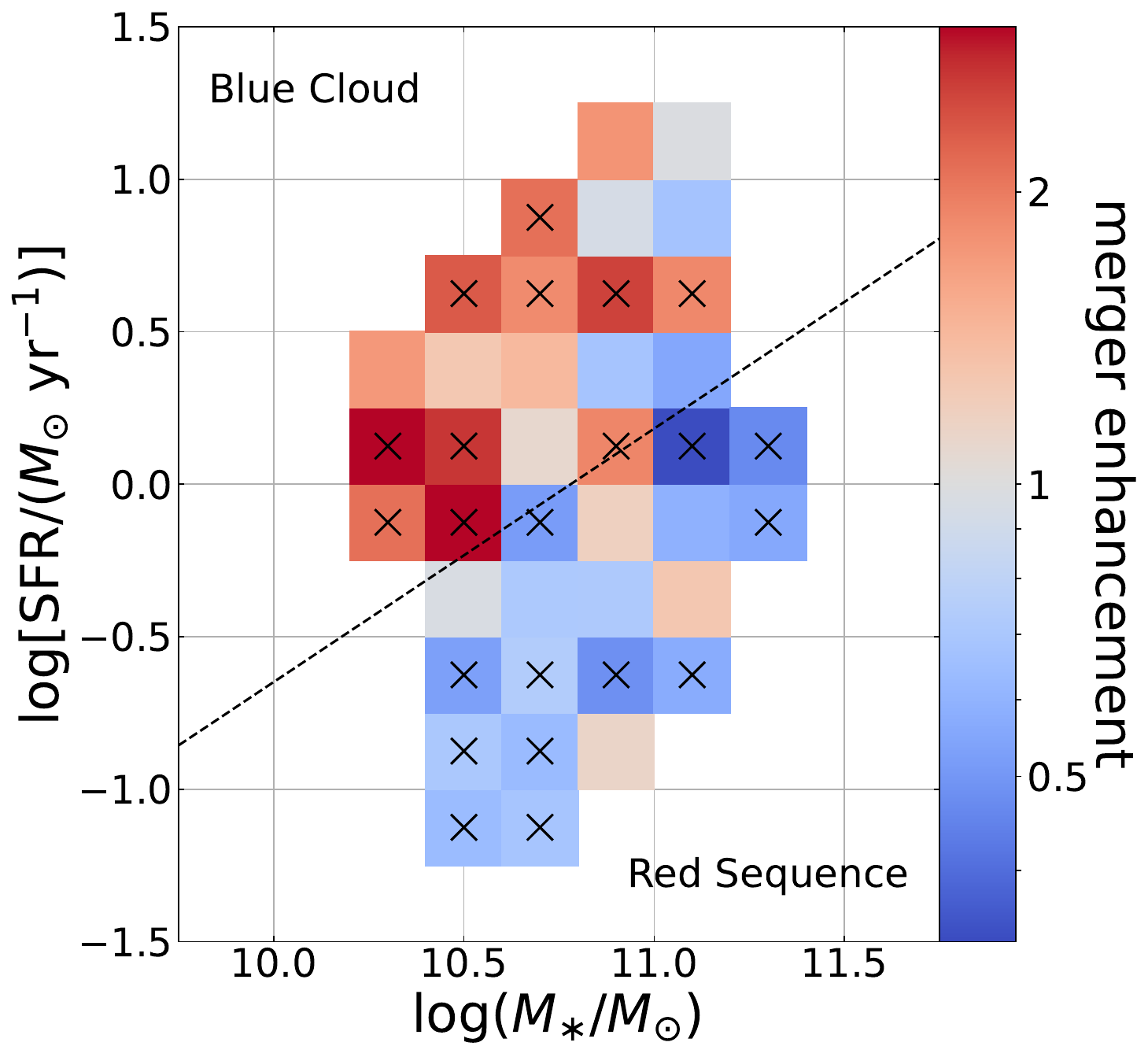}
  \caption{Merger enhancement $\langle f_{\rm mg,AGN}\rangle / \langle f_{\rm mg,control} \rangle$ of the sample in equally (logarithmically) spaced bins of stellar mass and star formation rate. Bins are only shown if they contain at least 50 AGN hosts and 50 control galaxies. Bins marked with a cross have $p({\rm  enh} > 1)$ or $1 - p({\rm enh} > 1) > 0.68$  (i.e. the bins where merger enhancement deviates from 1 by at least $1 \sigma$). The dashed line indicates our adopted divide between blue cloud and red sequence, determined by the parent galaxy population.}
  \label{fig:enh-mstar-sfr}
\end{figure}

Given that stellar mass and star formation rate are known to separate galaxies into the distinct populations of the star-forming blue cloud and quiescent red sequence, the trends with $M_{\ast}$ and sSFR suggest an underlying difference between AGN triggering in these two populations. Figure~\ref{fig:enh-mstar-sfr} investigates this by showing median merger enhancement, $f_{\rm mg,AGN} / f_{\rm mg,control}$, of our sample binned in equally-spaced bins in the $\log M_{\ast} - \log \mathrm{SFR}$  plane. The merger enhancement distributions compared in each bin are generated as described at the beginning of this section, and crosses mark bins where $p({\rm enh} > 1)$ or $1 - p({\rm enh} > 1) > 0.68$, i.e. where the enhancement differs from unity by at least $1\sigma$. Here we see a clear difference in the merger fractions measured among the two populations, with significant merger excesses among AGN hosts across the blue cloud and a significant lack thereof in parts of the red sequence. The crossover point between these behaviours roughly follows the location of the green valley.

To succinctly compare the differences in merger fractions in the blue cloud and red sequence, we divide the populations by estimating the location of the green valley in the $\log M_{\ast} - \log \mathrm{SFR}$ plane for our sample. The dashed black line in Figure~\ref{fig:enh-mstar-sfr} shows this by-eye division, defined as $\log \mathrm{SFR} = 0.83 \log M_{\ast} - 8.96$. Above the divide, in the blue cloud, we find that the median merger fraction among AGN is $f_{\mathrm{mg, AGN}} = 1.07_{-0.14}^{+0.19}$\% compared to that of controls, $f_{\mathrm{ mg, control} = 0.52_{-0.09}^{+0.14}}$\%, for a median enhancement of $2.03_{-0.63}^{+0.92}$. We find that $p({\rm enh} > 1) = 0.953$: the positive enhancement is significant at the $2 \sigma$ level. Conversely, in the red sequence, we see a deficit of mergers among AGN hosts, with $f_{\rm mg, AGN} = 3.15_{-0.27}^{+0.37}$\% and $f_{\rm mg, control} = 5.01_{-0.35}^{+0.45}$\% and a median enhancement of $0.65_{-0.09}^{+0.10}$. Here we find that $p({\rm enh} > 1) = 0.0041$, so this lack of mergers is highly significant. These results reflect the substantial decrease of detected merger fraction with sSFR seen in both AGN hosts and controls, but they also suggest a different relationship between AGN activity and galaxy merger status depending on the star formation properties and by extrapolation the cold gas fractions of the merging galaxies.

\subsubsection{\label{subsubsec:f-mg-bt}Merger fraction and bulge fraction}
In addition to stellar properties, we investigate trends of merger fraction with a simple measure of overall galaxy morphology, bulge-to-total ratios from the catalogue of~\citet{simard+11}. The result is shown in panel (c) of Figure~\ref{fig:f-mg-1d}. The correlation between $f_{\mathrm{mg}}$ and B/T is the strongest of any property we examined, with Pearson correlation coefficients of 0.95 (AGN; $p = 2.8 \times 10^{-5}$) and 0.98 (controls; $p = 1.8 \times 10^{-7}$). This shows that the more bulge-dominated a galaxy is, the more likely it is to be identified as a post-merger by our CNN classifier, regardless of whether it hosts an AGN. This in and of itself is hardly surprising, as it is generally believed that mergers reorganize stellar orbits into bulges, so one might expect that very few post-mergers are disc-dominated. However, this also brings up the concern that bulge dominance may itself be a classification criterion for the CNNs rather than simply correlating with presence of post-merger features. This possibility is further explored in Section~\ref{subsec:results-validation}.

\subsubsection{\label{subsubsec:f-mg-z}Merger fraction and redshift}

Due to cosmological surface brightness dimming and the fact that merger detection typically relies on identifying low-surface-brightness features, it may be expected for identified merger fractions to decline with redshift and a speculated merger enhancement among AGN hosts (or any specific subpopulation) to therefore be detected at a lower level or missed entirely~\citep{pierce+23}. Conversely, our finding in Section~\ref{subsubsec:f-mg-m-sfr} of a merger enhancement among the AGN hosts only in star-forming galaxies would predict an increase in overall enhancement with redshift, given that star formation on the whole increases with redshift until cosmic noon \citep[$z \sim 2$;][]{md14}. Panel (d) of Figure~\ref{fig:f-mg-1d} shows merger fractions of AGN hosts and controls in redshift bins. We observe negative correlations between $f_{\rm mg}$ and $z$ for both samples, though neither is particularly significant (AGN: $p = 0.081$; controls: $p = 0.018$), likely due to the flattening of predicted $f_{\rm mg}$ above $z \sim 0.1$. In fact, 84\% of the galaxies with a majority of votes for merger lie at $z < 0.1$.

Given that the bulk of the decline in observed merger fraction occurs between $z \sim 0$ and $z \sim 0.1$, corresponding to a surface brightness dimming of at most 0.2 mag, it seems unlikely that the redshift dependence of observed merger fraction is entirely due to cosmological dimming. However, the redshift-dependent mass- and $L$[\textsc{Oiii}]-completeness (see Figure~\ref{fig:agn-mz}) may play a role here, as merger fraction is seen to decline with stellar mass, particularly among AGN hosts (see panel (a) of Figure~\ref{fig:f-mg-1d} and Section~\ref{subsubsec:f-mg-m-sfr}), as well as the difference seen between AGN above and below $L\mathrm{[\textsc{Oiii}]}\sim 10^{40}$ erg s$^{-1}$ (see panel (e) and Section~\ref{subsubsec:f-mg-lum}). Given the narrow redshift range of this study, our results here cannot be extrapolated to higher redshift.

\subsubsection{\label{subsubsec:f-mg-lum}Merger fraction and AGN luminosity}

It has been suggested that the role of galaxy mergers in AGN triggering may differ depending on AGN luminosity and accretion rate: perhaps mergers are only needed to trigger the most luminous, rapidly accreting AGN~\citep{hh09, hkb14}. This study is not well poised to make statements on relationships with luminosity as our AGN are all at the lower end of the luminosity function, but we nevertheless examine any possible trends within our luminosity range in panel (e) of Figure~\ref{fig:f-mg-1d}. We observe a marginally-significant negative correlation between merger fraction and $L$[\textsc{Oiii}] ($r = -0.65;\ p = 0.043$), though the main source of the trend comes from the three lowest-luminosity bins, which are below the AGN luminosity cuts used by many studies. In fact, if anything, the slight uptick around $L\mathrm{[\textsc{Oiii}]}\sim 10^{41}$ erg s$^{-1}$ could be suggestive of a trend to higher luminosities.

\subsection{\label{subsec:results-validation}CNN validation and comparison with visual inspection}

While it is a strength of CNNs that they learn what features to extract at training time, this quality can lead to difficulty telling which aspects of an image lead to the CNN's classification. Are our CNNs picking up on specific tidal features indicative of recent mergers, or are there other properties shared by the post-mergers in the training set, such as being more bulge-dominated, driving the classifications?

To get a more intuitive sense of the CNNs' selection criteria, as well as compare the CNN to human classifiers, two of the authors (MSAM and CV) performed visual inspection on the 70 SDSS galaxies representing the most confident CNN-selected mergers $(P_{\rm mg} > 0.5$ for $\geq 84\%$ of CNNs; 24 AGN hosts and 46 controls). We matched these to confident nonmergers in $M_{\ast}$, SFR, and AGN classification to create a balanced sample for inspection. While these small subsamples may not be fully representative of the scope of CNN-classified mergers and nonmergers, they should give us a sense of whether or not the CNN-selected mergers are visually identifiable. The galaxies were inspected blind to their merger classification and AGN status, although the human classifiers were aware of the 50/50 merger/nonmerger split. 

The human classifiers inspected each image using the Zooniverse\footnote{\url{Zooniverse.org}} interface, with a $2\times2$ grid showing views of the image with two different saturation levels and both the original and CNN-rescaled pixel sizes. Classifiers were first asked to identify each galaxy's morphology as bulge-dominated, disc-dominated, or unable to tell, as well as if they saw any merger features, which had to be specified from a list (asymmetry, tidal tail, shell, multiple cores, or interacting companion). Classifiers were additionally asked whether they saw any non-interacting galaxies in the field. Finally, comments were used to indicate uncertainty in their merger classifications (e.g. `mild asymmetry' or `possibly a spiral arm'). These were used to separate the visually-classified mergers into `possible' and `certain' subcategories.

\begin{figure}
  \centering
  \includegraphics[width=6cm]{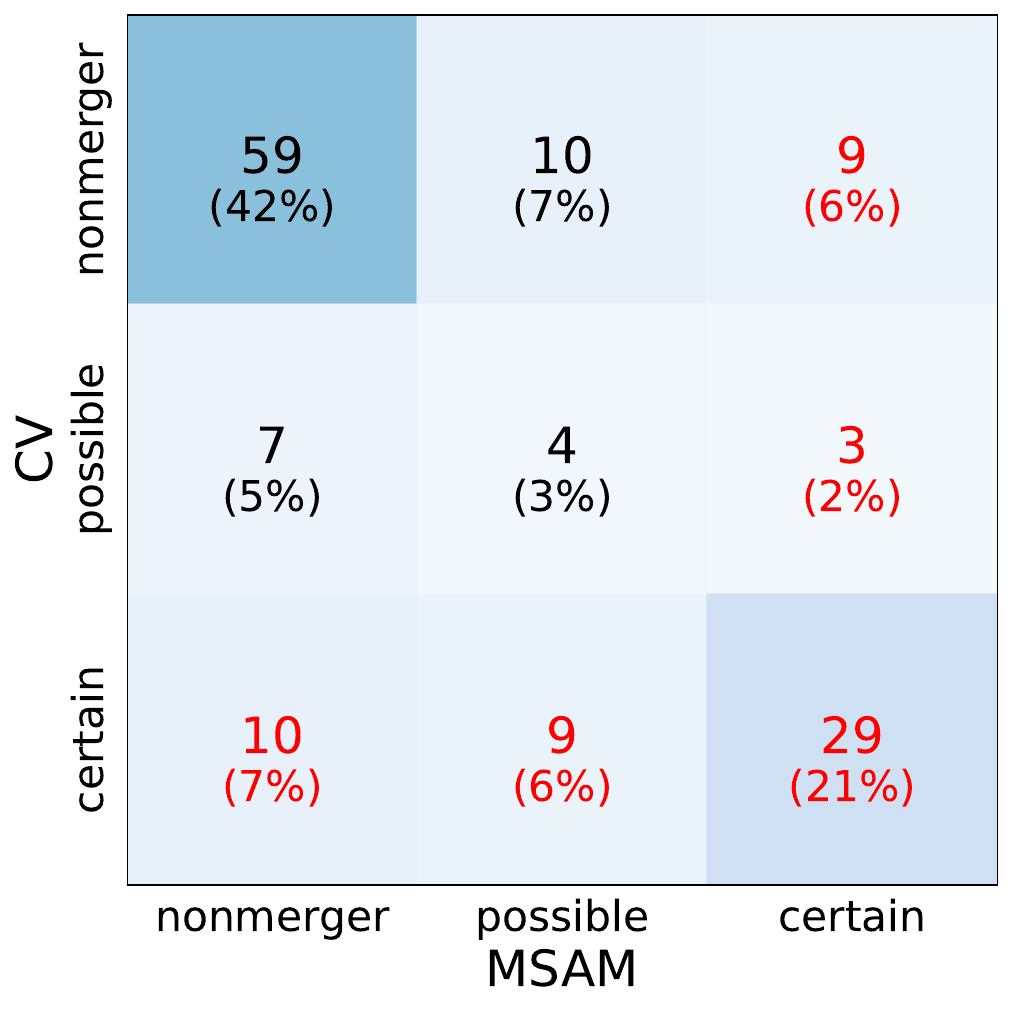}
  \caption{Confusion matrix showing agreement between the human visual classifiers, including both possible and certain mergers. Red text indicates classifications considered as visual mergers, while black text indicates nonmergers.}
  \label{fig:cm-humans}
\end{figure}

 The classifiers agreed with each other in 74\% of cases, with 42\% agreed as nonmergers, 21\% as certain mergers, and an additional 11\% as at least possible mergers. We define our visual merger classification as galaxies labelled as a certain merger by least one classifier (including those where the other classification was nonmerger). Figure~\ref{fig:cm-humans} shows the agreement between the visual classifiers, with the collective `MSAM or CV' classification indicated by text colour.

\begin{figure*}
    \centering
    \includegraphics[width=12cm]{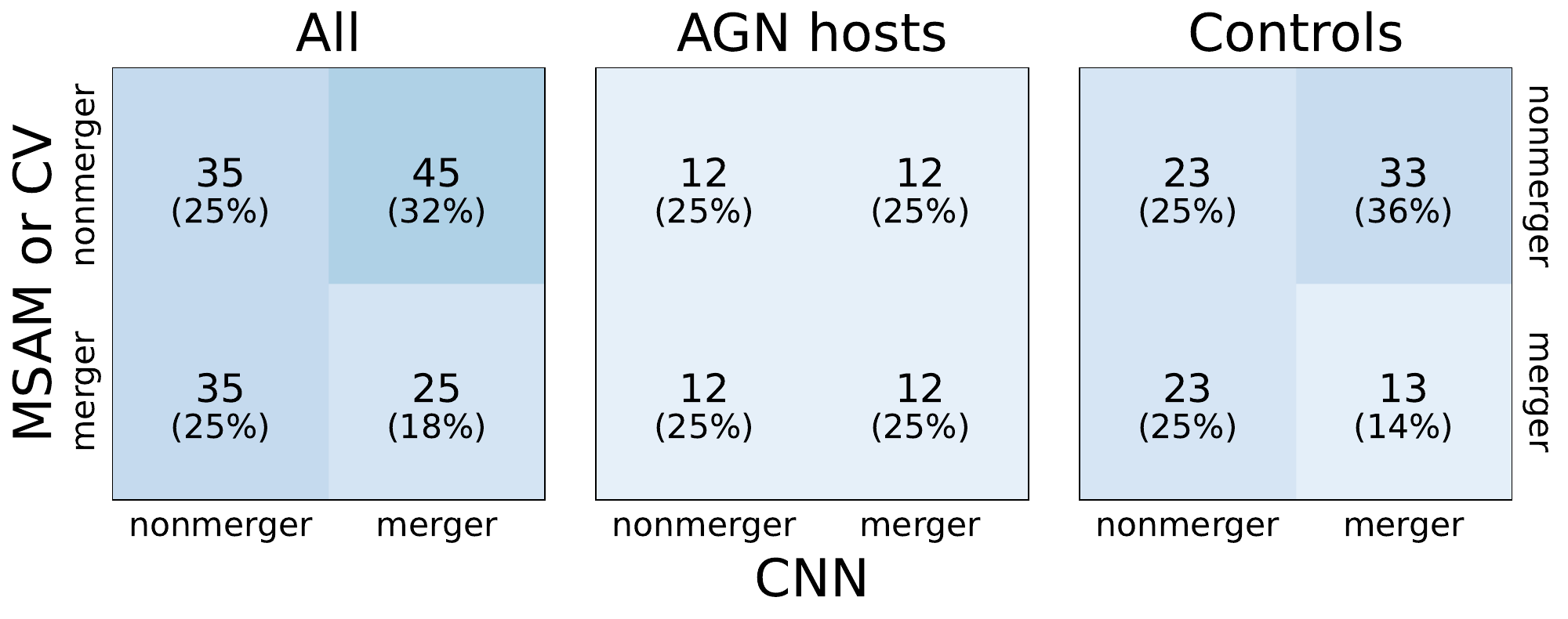}
    \caption{Confusion matrices showing (dis)agreement between the combined visual classifications (`MSAM or CV') and the CNN classifications, over all galaxies inspected (left), AGN hosts only (centre), and control galaxies only (right).}
    \label{fig:cm-summary}
  \end{figure*}

Our visual classifications agree with the CNN classifications in only 43\% of cases. As shown in Figure~\ref{fig:cm-summary}, MSAM and CV collectively classified the CNN nonmerger set as being $50 \pm 5.9 \%$ mergers, while they only identified visual merger features in $35.7_{-5.2}^{+6.0}\%$ of the CNN mergers (errors given by beta distribution 1$\sigma$ intervals). Overall, there appears to be very little similarity between the CNN-identified mergers and what human visual classifiers deem as likely mergers.
  
Figure~\ref{fig:sdss-ex} gives examples of the visually classified images and hence intuitively demonstrates where the CNN and humans diverge in merger classification criteria. While several of the CNN-predicted mergers show strong tidal features and high levels of disturbance, many appear smooth and featureless. We would expect high nonmerger contamination among these due to the rarity of mergers in the parent sample. Conversely, many of the CNN-predicted nonmergers show features, particularly spiral arms, that could lead to merger flags based on our visual criteria, but may well be due to other processes than mergers. Coupled with the observed positive correlation between predicted merger fraction and $B/T$ (see Figure~\ref{fig:f-mg-1d}c), as well as the tendency of the CNNs to predict more false positives among galaxies with higher mass ratios of and less time since their most recent merger (even over timescales of Gyr), this points to the CNNs identifying mergers by overall galaxy shape as well as a few very specific features, but asymmetry on its own is not enough to merit a high merger probability. This interpretation also provides an explanation for why the human classifiers found a lower fraction of visual mergers among the CNN mergers in the control galaxies compared with the AGN hosts (see the right columns of the central and right panels of Figure~\ref{fig:cm-summary}), since the control sample contains proportionately more of these bulge-dominated galaxies. In summary, while visual classification shows poor agreement with the CNN classification, this can be explained by the differing features identified by humans and the CNN.

  To better understand how well human classifiers can identify post-mergers from images, MSAM and CV also inspected a subset of the TNG training images. They were shown mock images of 46 post-mergers and 46 matched nonmergers, sampled across the $M_{\ast}$ and $z$ distributions, with the knowledge that the dataset was evenly split. They viewed the idealised and realistic images simultaneously and were asked which image, if either, showed merger features. 

  Compared with the true merger state from the simulation, MSAM/CV were unable to consistently identify actual post-mergers, each correctly identifying merger status of 53/66\% of the idealised images and 49/58\% of the realistic images. Both classifications from CV are consistent with random guessing, while MSAM achieved slightly higher accuracy at the cost of labelling many (29\% for idealised and 36\% for realistic) false positives. Both classifiers saw a slight but not significant increase in accuracy comparing realistic to idealised images, indicating that even in the idealised stellar maps with no noise or sky added, mergers are not visually distinguishable. However, we do observe a slight increase of true positive rate with decreasing time since the merger (no trend was observed with mass ratio). Evidently post-mergers in TNG100 do not all exhibit the features used to visually classify galaxies as merging, though this appears more likely to be the case immediately after coalescence.

  The inconsistency between visual and CNN classification raises the question: do galaxy mergers in cosmological simulations actually look like real galaxy mergers? Several previous works have assessed the suitability of TNG for comparisons to SDSS (also using \textsc{RealSim}), though they do not focus specifically on mergers. \citet{hc+19} showed that TNG100 produces a population of galaxies whose morphological properties broadly agree with SDSS at $z \sim 0.05$, spanning the full range of morphological types and reproducing global relations such as the size-mass relation. However,~\citet{zanisi+21} found that a neural network could identify differences between the populations of galaxies in TNG compared with SDSS, calculating a distance metric between the simulated and real galaxy populations. Notably, when split into star-forming and quiescent samples, the simulated star-forming population displayed a smaller distance to its SDSS counterpart than the quiescent population did, indicating that TNG reproduces more realistic star-forming galaxies than quiescent. \citet{eisert+23} sought to connect image realism with specific galaxy properties, using deep learning to compare TNG galaxies with those observed by HSC They found that overall, $\sim 67\%$ of TNG100 galaxies reside within the domain of visual appearances spanned by observations, while the other 33\% are `out of domain,' which can be interpreted as realistic/unrealistic. They connected a high out-of-domain score (implying poor realism) with larger radii and asymmetry, higher concentration, and lower ellipticity (see their Figure 12), some of which are features associated with red sequence galaxies, though they did not find similar trends with galaxy colour, $B/T$, or $M_{\ast}$. We note that we observe a decline of CNN performance with increasing physical radius but not angular size (see Figure~\ref{fig:fpr-tpr}cd), though this may be easily explained by the decrease in number of galaxies above a certain size.
  
To summarise, visual inspection on subsets of both our science sample and our training sample has shown that while many of the CNN-selected mergers show no visually obvious signatures, neither do many of the actual post-mergers found in TNG. Conversely, while many of the CNN-selected nonmergers show asymmetry and other features suggestive of a merger, so do TNG galaxies that have not undergone a merger in several Gyr. While recent studies in the field suggest that a substantial fraction of our training sample may be `unrealistic' in terms of their resemblance to real galaxies, we cannot say for sure whether the discrepancy between simulated mergers and what humans identify as mergers is the result of poor simulation realism or an indication that visual classification is not as accurate at identifying true post-mergers as previously thought.

\section{Discussion}
\label{sec:discussion}

Overall, we find no enhancement in merger fractions between Seyfert 2 hosts and mass- and redshift-matched control galaxies ($f_{\rm mg, AGN} = 2.19_{-0.17}^{+0.21}$\% vs $f_{\rm mg, control} = 2.96_{-0.20}^{+0.26}$\%), indicating that mergers are not the dominant cause of most supermassive black hole accretion in obscured, low-luminosity AGN in the nearby Universe, and secular processes must play an important role. However, when splitting our sample into star-forming and quiescent populations, we find a significant merger excess among AGN hosts in the blue cloud compared with controls, while there are significantly fewer mergers among AGN hosts than controls in the red sequence. This is seen when comparing merger fractions of AGN hosts with controls in bins of stellar mass and specific star formation rate (Figure~\ref{fig:f-mg-1d} panels (a) and (b)), as excesses in merger fraction of AGN hosts over controls are seen at low $M_{\ast}$ and high sSFR, and the divide becomes clear when visualising merger enhancements over the $M_{\ast} - \mathrm{SFR}$ plane (Figure~\ref{fig:enh-mstar-sfr}).

We note that due to the fact that major mergers are rare in the general population, that even with high detection accuracy as achieved here, the final sample of mergers is expected to have low purity. This is due to the fact that the false positives on the much larger non merger sample will be higher in number than the true positives on the much smaller number of major mergers. This also results in the excess of mergers being underestimated. For example, assuming the accuracy achieved here and a fraction of mergers in the full sample of 5\% (as observed here), the true excess would be underestimated by about a factor of $\approx$2. This means that the actual excess seen in the star-forming galaxies is likely higher, while the upper limit on the excess in the full sample is still consistent with an enhancement of mergers in the AGN sample of a factor of $\sim$ 2.

This result suggests that the ability of a merger to trigger an AGN depends on the cold gas content of the galaxy, as higher sSFR in the post-merger indicates the presence of cold gas. If there is no gas for a merger to drive to the centre, then the merger has no positive effect on the likelihood of an AGN to form, as we actually observe a relative (low-significance) lack of mergers among AGN hosts compared with controls in non-star-forming galaxies. 
This is broadly consistent with the conclusions of~\citet{sbh15}, though our data and results differ somewhat. Studying a sample drawn from the same MPA-JHU catalogue, \citet{sbh15} found a correlation between AGN fraction and host galaxy interaction rate, but this correlation disappeared when controlled for \emph{central} star formation. They concluded that the key requirement for AGN formation is availability of cold gas at the centre of the galaxy. This work finds an excess among AGN hosts only in galaxies with higher specific star formation \emph{over the entire galaxy}, suggesting that mergers are a vehicle for moving this gas inwards to the centre, when it is present in the outer regions.

Unlike~\citet{ellison+19}, who studied a similar sample of optically-selected low-luminosity AGN hosts from SDSS, we find no significant correlation between [\textsc{Oiii}] luminosity and merger fraction. Our observation of a general decrease of $f_{\mathrm{mg}}$ with stellar mass also disagrees with their finding of an increase of $f_{\mathrm{mg}}$ above $M_{\ast} \sim 10^{10.6}\ M_{\odot}$. However, the complete lack of agreement between our CNNs and expert human classifiers on our own galaxies (see Section~\ref{subsec:results-validation}) indicates that the CNN classifications are probing a different population of galaxies from visual classifications. \citet{ellison+19} based their merger classifications on visual inspection and included systems identified as both pre- and post-coalescence in their merger correlation calculations, compared with our CNN classifier only looking for post-mergers, so it is likely that our two studies probe very different populations of galaxies both identified as `mergers.'

\citet{ellison+19} additionally report much higher merger fractions in both their AGN hosts and control galaxies than what we find here: $\sim$18\% of their total sample are identified as post-mergers, compared to our $\sim$4\%. As they base their merger identification on imaging from CFIS, which they demonstrate to be deeper and of higher quality than SDSS imaging, it is likely that our images are missing low-surface-brightness features identifiable in CFIS images of the same galaxies. Further, as our CNN classifier has been demonstrated to identify galaxies as post-mergers without the presence of human-identifiable merger signatures, the converse may be true as well, as our visual inspection experiments identified a number of galaxies with asymmetries that none the less had confident nonmerger classifications by the CNNs.

As star formation has generally declined from $z \sim 2$ to $z \sim 0$~\citep{md14}, our observation of a merger enhancement only among AGN hosts in the star-forming blue cloud suggests that the merger-AGN connection should be stronger at higher redshift, consistent with the models of~\citet{db12} and~\citet{hkb14}. Our redshift window of $z \sim 0 - 0.3$ is too narrow to see sSFR increase appreciably with $z$ in our sample, and we also see no increase of merger enhancement: in fact, we see the opposite, likely affected by both surface brightness dimming and the redshift-dependent mass and luminosity completeness of SDSS. While most higher-redshift studies of the merger-AGN relationship have focused on high-luminosity quasars,~\citet{cisternas+11} looked at host galaxy morphologies of AGN with $L_{\mathrm{X}}\ (2-10\ \mathrm{keV}) \lesssim 10^{45}$ erg/s at $z \sim 0.3 - 1$ in the COSMOS survey, which represent a similar luminosity range to ours. They found no overall correlation between merger features and AGN activity. They further found no significant dependence of merger fraction on stellar mass within the AGN population, though they did not comment on star formation properties or gas fractions in their galaxies. For both their AGN hosts and their controls, they found $34-35$\% to be bulge-dominated and $65-66$\% to be disc-dominated, compared to our $74$\% of AGN hosts and $69$\% of controls being disc-dominated ($B/T < 0.5$). Their observation of no overall enhancement is thus consistent with ours, though we cannot compare specifically between the blue cloud and red sequence populations. It is worth noting that~\citet{cisternas+11} also found substantially higher merger fractions than this work did, with $15$\% of their AGN hosts and $13$\% of their controls being classified with strong distortions. As COSMOS is a much deeper survey than SDSS, it is difficult to say with certainty whether this difference is related to the increased redshift and implied increased gas fractions and SFRs or simply due to the increased sensitivity (though it may be noted that their observed merger fraction is consistent with the post-merger fraction found by~\citealt{ellison+19} at $z \sim 0$).

Our results are also broadly compatible with theoretical predictions. In both the \textsc{eagle} and Magneticum Pathfinder simulations, mergers are found not to dominate black hole fuelling, but they are most relevant at the highest luminosities ($L_{\mathrm{bol}} \gtrsim 10^{46}$ erg s$^{-1}$), far outside the region examined in this work~\citep{mcalpine+20, steinborn+18}. Both observe a decreasing merger enhancement with stellar mass, with fewer mergers seen among AGN hosts than control galaxies at $M_{\ast} \gtrsim 10^{11} M_{\odot}$. This is similar to our results in panel (a) of {Figure}~\ref{fig:f-mg-1d}, though we see the transition at the lower mass of $M_{\ast} \sim 10^{10.5} M_{\odot}$.

In more general terms, hydrodynamical simulations have shown that galaxies with AGN are often characterized by a larger gas density within the resolved accretion region around the central SMBH~\citep[e.g.][Figure 9]{steinborn+18}. Whilst in all galaxies these central gas densities decrease with cosmic time, particularly following the overall galactic cosmic starvation, galaxies with evident AGN activity and larger SMBH masses have a tendency for higher central densities, a trend largely independent of their merger histories but possibly more related to the conditions of the larger scale gas distribution. The same simulations also show that mergers tend to trigger more AGN activity, but their frequency among the AGN population remains limited to $<20\%$, pointing to a contained role of mergers in boosting AGN. This also matches our results of relatively low merger fractions. The simulations also clearly predict that the AGN are much more common in star forming galaxies, pointing to a close and statistically sound correlation between AGN activity and SFR, although the models also predict AGN fractions independent of the AGN luminosity, matching our results of a higher merger enhancement in star-forming galaxies (see Figure \ref{fig:enh-mstar-sfr}). 

  These theoretical predictions are in broad agreement with our observations that suggest a relatively small difference in the merger fractions of active and inactive galaxies and a tendency for galaxies with larger sSFR to have more AGN activity. More recent studies~\citep{smethurst+23} conducted on the AGN-Horizon simulation have shown that the BH mass-galaxy mass scaling relation is preserved in all types of galaxies irrespective of their bulge-to-total ratio and level of merger activity, even in galaxies with almost quiescent assembly histories, further supporting the idea of a loose link between mergers and AGN activity. Simulations with Lagrangian hyper-refinement~\citep{aa+21}, however, have also shown that although sub-pc inflow rates do correlate with nuclear star formation, they might decouple with the larger scale SFR in the host galaxy. Therefore, our results are consistent with a picture in which the connection between mergers and AGN activity depends on the galaxy's gas fraction (see Figure \ref{fig:enh-mstar-sfr}).

\section{Conclusions}
\label{sec:conclusions}

The relative importance of galaxy interactions to the fuelling of supermassive black holes has long been a subject of debate. While major mergers have been shown to trigger the necessary gas inflows, observational evidence remains inconclusive due to the difficulty of consistently identifying merger features, particularly in galaxies containing AGN.

In this paper, we have approached this problem by using deep learning techniques to detect galaxy mergers in a sample of $\sim$8500 Type 2 Seyferts at $z < 0.3$ compared with mass- and redshift-matched inactive control galaxies. We have accomplished this by using supervised learning with an ensemble of convolutional neural networks trained to identify post-mergers in the IllustrisTNG simulation, based on images processed to mimic SDSS $gri$ observations. Comparing identified merger fractions in our two samples, we find the following:
\begin{itemize}
\item There is no significant merger enhancement among low-redshift Seyfert 2 galaxies compared with inactive galaxies at the same stellar mass and redshift, with our CNN ensemble finding $f_{\rm mg, AGN} = 2.19_{-0.17}^{+0.21}$\% and $f_{\rm mg, control} = 2.96_{-0.20}^{+0.26}$\%. This indicates that galaxy mergers are not the dominant trigger of low luminosity obscured AGN in the nearby Universe.
\item The fraction of mergers among AGN hosts decreases with stellar mass, while it remains constant for controls. AGN hosts with $M_{\ast} \lesssim 10^{10.5} M_{\odot}$ are more likely to have undergone a recent merger than control galaxies at the same mass, while those with $M_{\ast} \gtrsim 10^{10.5} M_{\odot}$ are less likely.
\item Merger fraction for both AGN hosts and controls decreases with specific star formation rate and increases with bulge-to-total fraction. The decrease with sSFR is less pronounced for AGN hosts than for controls, and a merger excess is observed at sSFRs above $\sim$10$^{-10.5}$ yr$^{-1}$.
\item When separated in the $M_{\ast}$-SFR plane (Figure~\ref{fig:enh-mstar-sfr}), there is a significant difference in merger activity of AGN hosts relative to controls depending on the stellar populations of the galaxies. We observe, relative to controls, both
  \begin{enumerate}
  \item a significant merger enhancement of $2.03_{-0.63}^{+0.92}$ among AGN hosts in the blue cloud ($f_{\rm mg, AGN} = 1.07_{-0.14}^{+0.19}$\% and $f_{\rm mg, control} = 0.52_{-0.09}^{+0.14}$\%; $p({\rm enh} > 1) = 0.953$) and
  \item a significant lack of mergers (${\rm enh} = 0.65_{-0.09}^{+0.10}$) among AGN hosts in the red sequence ($f_{\rm mg, AGN} = 3.15_{-0.27}^{+0.37}$\% and $f_{\rm mg, control} = 5.01_{-0.35}^{+0.45}$\%; $p({\rm enh} > 1) = 0.0041$),
  \end{enumerate}
  suggesting that major mergers have very different impacts on black hole accretion depending on the specific star formation rate, and by implication cold gas fraction, of the host galaxies involved. Mergers appear to have a positive impact on AGN formation in star-forming, gas-rich galaxies, helping to drive gas from the disc to the centre. Conversely, mergers have no significant effect on AGN in galaxies that are overall gas-poor.
\item Convolutional neural networks trained on simulated galaxy mergers agree very little with human classifiers when examining the same set of observations. Human classifiers are found to perform very poorly at identifying simulated post-merger galaxies, suggesting that these galaxies actually look very different from the human notion of a galaxy merger and that deep-learning approaches will be more reliable identifiers of post-mergers going forward.
\end{itemize}

\section*{Acknowledgements}


We thank the anonymous referee for their comments, which improved the quality of this paper.

This project has received funding from the European Union's Horizon 2020 research and innovation programme under the Marie Sk{\l}odowska-Curie grant agreement No 860744. Computing time support provided by Royal Society research grant RGS/R1/231499. This work made use of Astropy:\footnote{http://www.astropy.org} a community-developed core Python package and an ecosystem of tools and resources for astronomy~\citep{astropy:2013, astropy:2018, astropy:2022}. This publication uses data generated via the \url{Zooniverse.org} platform, development of which is funded by generous support, including a Global Impact Award from Google, and by a grant from the Alfred P. Sloan Foundation. This research has made use of the VizieR catalogue access tools, CDS, Strasbourg, France (DOI: 10.26093/cds/vizier). The original description of the VizieR service was published in 2000, A\&AS 143, 23. AL is partly supported by the PRIN MIUR 2017 prot. 20173ML3WW 002 `Opening the ALMA window on the cosmic evolution of gas, stars, and massive black holes.' 

Funding for the SDSS and SDSS-II has been provided by the Alfred P. Sloan Foundation, the Participating Institutions, the National Science Foundation, the U.S. Department of Energy, the National Aeronautics and Space Administration, the Japanese Monbukagakusho, the Max Planck Society, and the Higher Education Funding Council for England. The SDSS Web Site is \url{http://www.sdss.org/}.

The SDSS is managed by the Astrophysical Research Consortium for the Participating Institutions. The Participating Institutions are the American Museum of Natural History, Astrophysical Institute Potsdam, University of Basel, University of Cambridge, Case Western Reserve University, University of Chicago, Drexel University, Fermilab, the Institute for Advanced Study, the Japan Participation Group, Johns Hopkins University, the Joint Institute for Nuclear Astrophysics, the Kavli Institute for Particle Astrophysics and Cosmology, the Korean Scientist Group, the Chinese Academy of Sciences (LAMOST), Los Alamos National Laboratory, the Max-Planck-Institute for Astronomy (MPIA), the Max-Planck-Institute for Astrophysics (MPA), New Mexico State University, Ohio State University, University of Pittsburgh, University of Portsmouth, Princeton University, the United States Naval Observatory, and the University of Washington.

\section*{Data Availability}


All data used in this study are publically available online. SDSS line data (including the MPA-JHU catalogue) were queried from the SDSS Catalog Archive Server (\url{cas.sdss.org/dr7}), while the fields used for cutout creation and observational realism come from the Data Archive Server (\url{das.sdss.org}). All IllustrisTNG catalogues and mock imaging come from their database at \url{tng-project.org/data}. The bulge-to-total ratios used are available from VizieR (DOI: 10.26093/cds/vizier.21960011).

On publication, the catalogue of merger predictions for SDSS galaxies will be published on VizieR, and scripts used to generate the predictions and results will be made available at \url{github.com/mathildaam}.



\bibliographystyle{mnras}
\bibliography{bibliography} 




\appendix

\section{CNN performance on TNG galaxies}
\label{appendix:perf}

  This section provides additional details on the performance of the CNNs on TNG training data, illustrating the stability of merger predictions with regard to changes in classification threshold probability as well as the possibility of merger prediction bias with different galaxy properties.

\begin{figure}
    \centering
    \includegraphics[width=8cm]{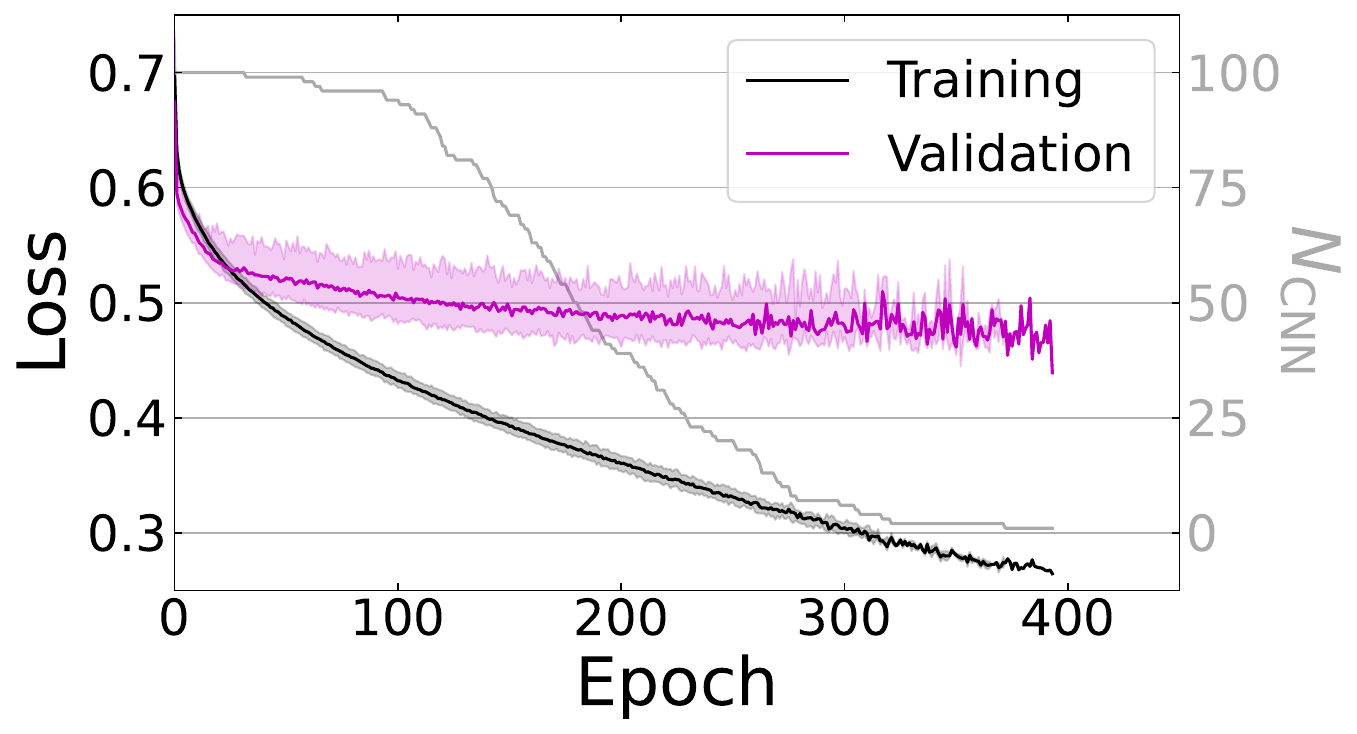}
    \caption{Loss function of the CNNs over the training period. Solid lines show median loss over all CNNs at each epoch, while shaded regions show 90\% of the CNNs. As training length is variable, the number of CNNs still training at a given epoch is shown in grey.}
    \label{fig:histories}
  \end{figure}

  Figure~\ref{fig:histories} shows the aggregated training histories of the CNN ensemble. While the loss function continuously improves for the training galaxies, it levels off quickly for the validation galaxies. Training is cut off after 50 epochs of no improvement to the validation loss, thereby limiting the amount of overfitting.

\begin{figure}
  \centering
  \includegraphics[width=8cm]{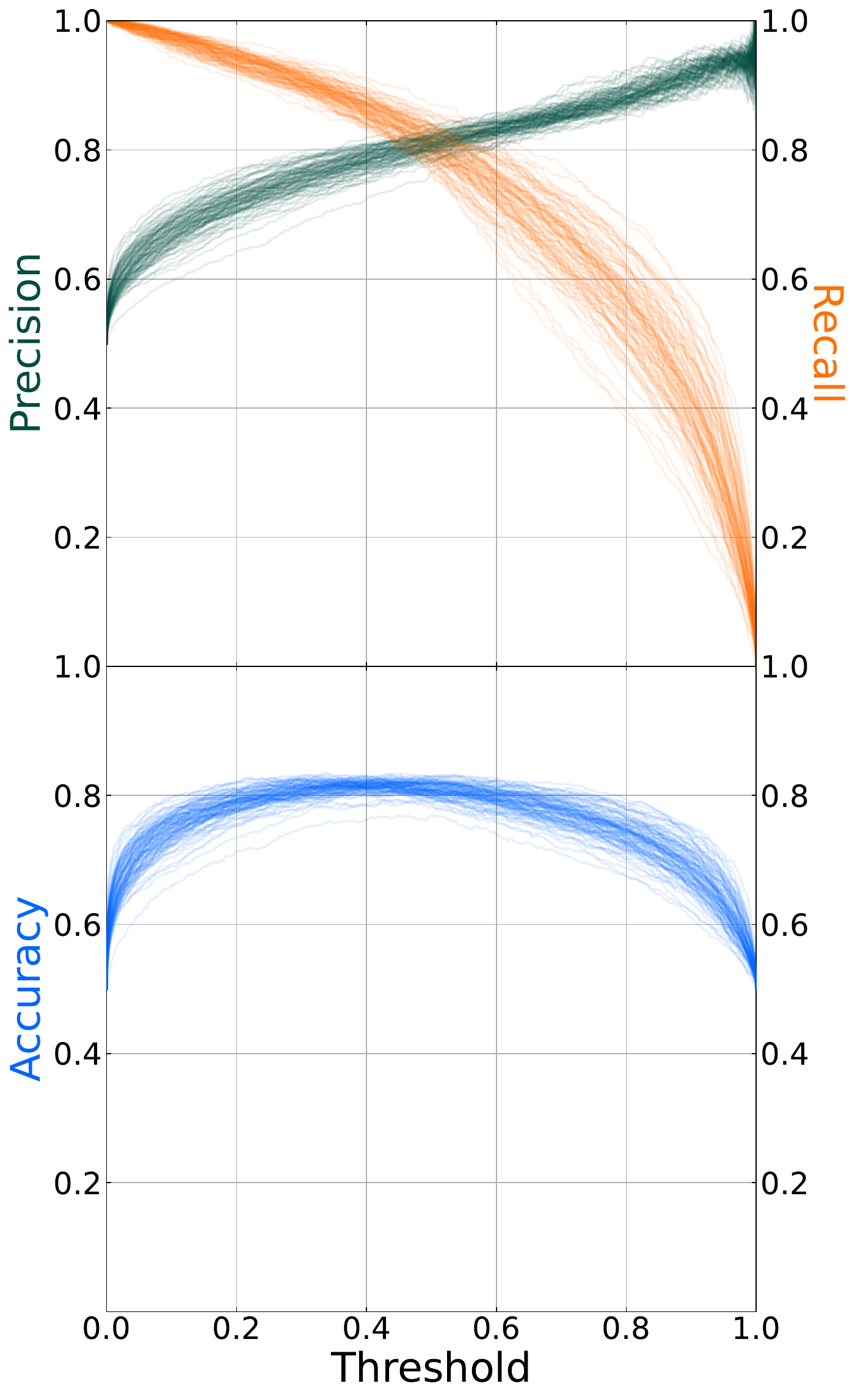}
  \caption{Precision and recall (top panel) and overall accuracy (bottom panel) as a function of threshold probability for our ensemble of 100 CNNs applied to the testing set of our IllustrisTNG training data. Each line represents a single CNN.}
  \label{fig:metrics}
\end{figure}

\begin{figure}
  \centering
  \includegraphics[width=8cm]{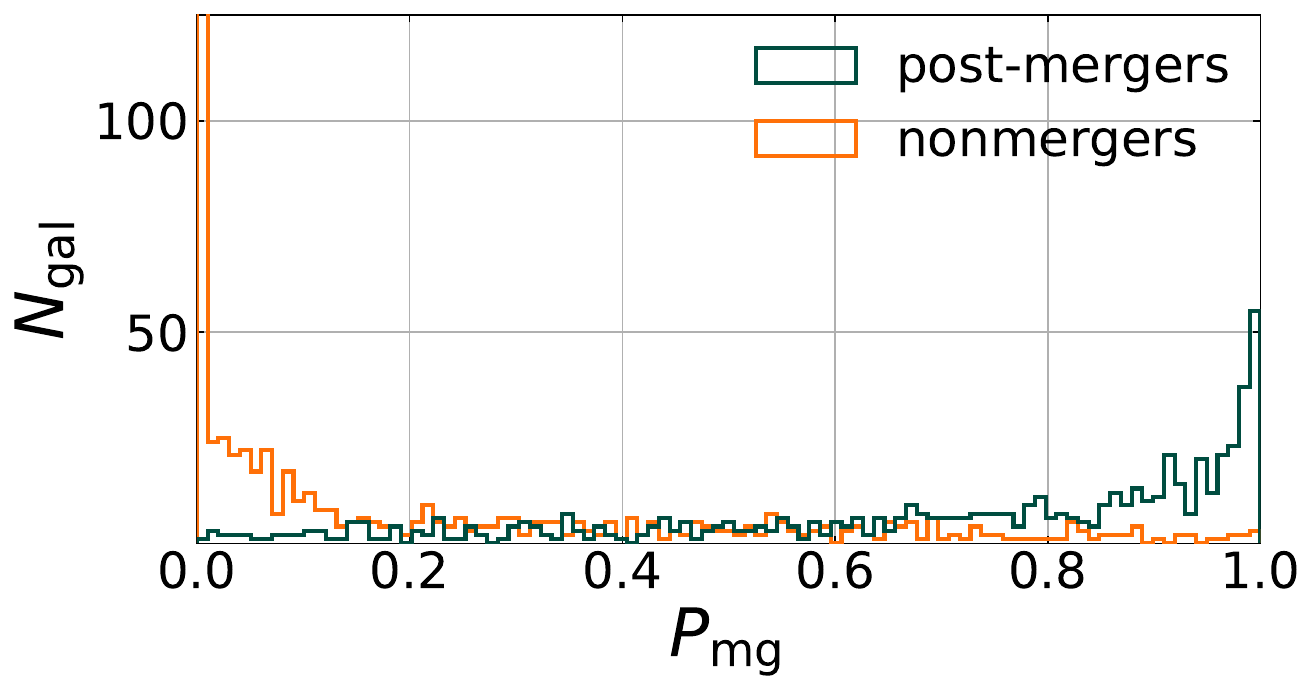}
  \caption{Predicted merger probabilities for the TNG testing set for one of our CNNs. The relative lack of highly confident post-mergers vs with highly confident nonmergers is typical, potentially due to the wider variation in appearance of post-mergers compared with nonmergers.}
  \label{fig:p-mg-ex}
\end{figure}

  Figures~\ref{fig:metrics} and~\ref{fig:p-mg-ex} illustrate the effect of changing the threshold between merger and nonmerger classification. Precision and recall (also called purity and completeness as in Section~\ref{subsec:results-cnn-training}) measure the fraction of correctly-identified galaxies out of the set of identified and actual positives, respectively (i.e. precision=TP/(TP+FP) and recall=TP/(TP+FN)). These vary with classification threshold in a complementary way, as shown in the top panel of Figure~\ref{fig:metrics}. Our adopted threshold of 0.5 is seen to balance both metrics. Overall CNN accuracy, seen in the bottom panel, is largely insensitive to prediction threshold within the central third of its range. This is due to the strong bimodality in $P_{\rm mg}$ predictions for most of our CNNs, an example of which is shown in Figure~\ref{fig:p-mg-ex}. For our balanced testing set, small changes in prediction threshold see more false positives balanced out by more true positives, or vice versa. For the observational dataset in which mergers are rare events, the tradeoff becomes more extreme, where changing the threshold to include one more true merger would be expected to add 10-20 false positives. However, we note that the trends seen in Figure~\ref{fig:f-mg-1d} persist over different choices of classification threshold, with the overall normalisation being the only change.

\begin{figure*}
  \centering
  \includegraphics[width=8cm]{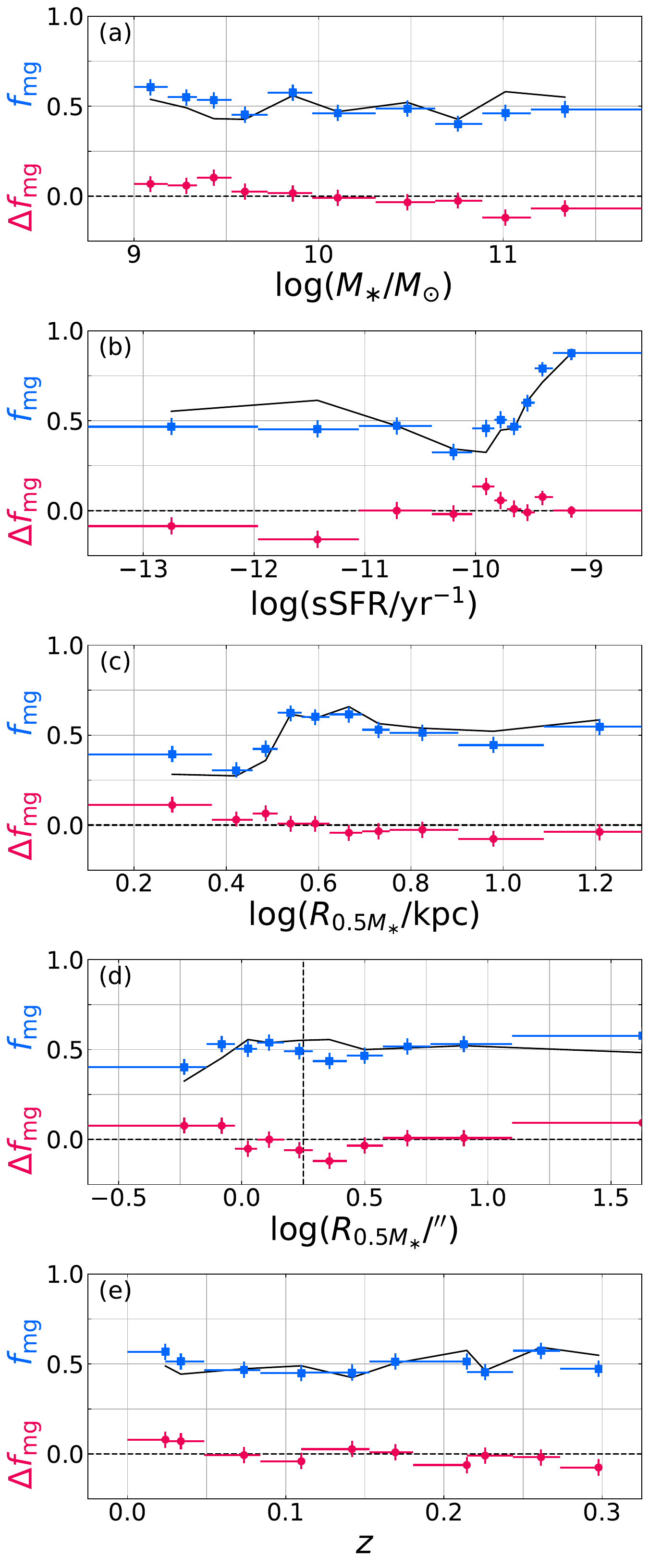}
  \caption{Predicted merger fraction of TNG training galaxies (blue squares) compared to actual merger fraction (black line) binned by (a) stellar mass, (b) specific star formation rate, (c) physical stellar half-mass radius, (d) angular stellar half-mass radius, and (e) redshift. Pink circles show residuals.}
  \label{fig:fmg-tng}
\end{figure*}

\begin{figure*}
  \centering
  \includegraphics[width=14cm]{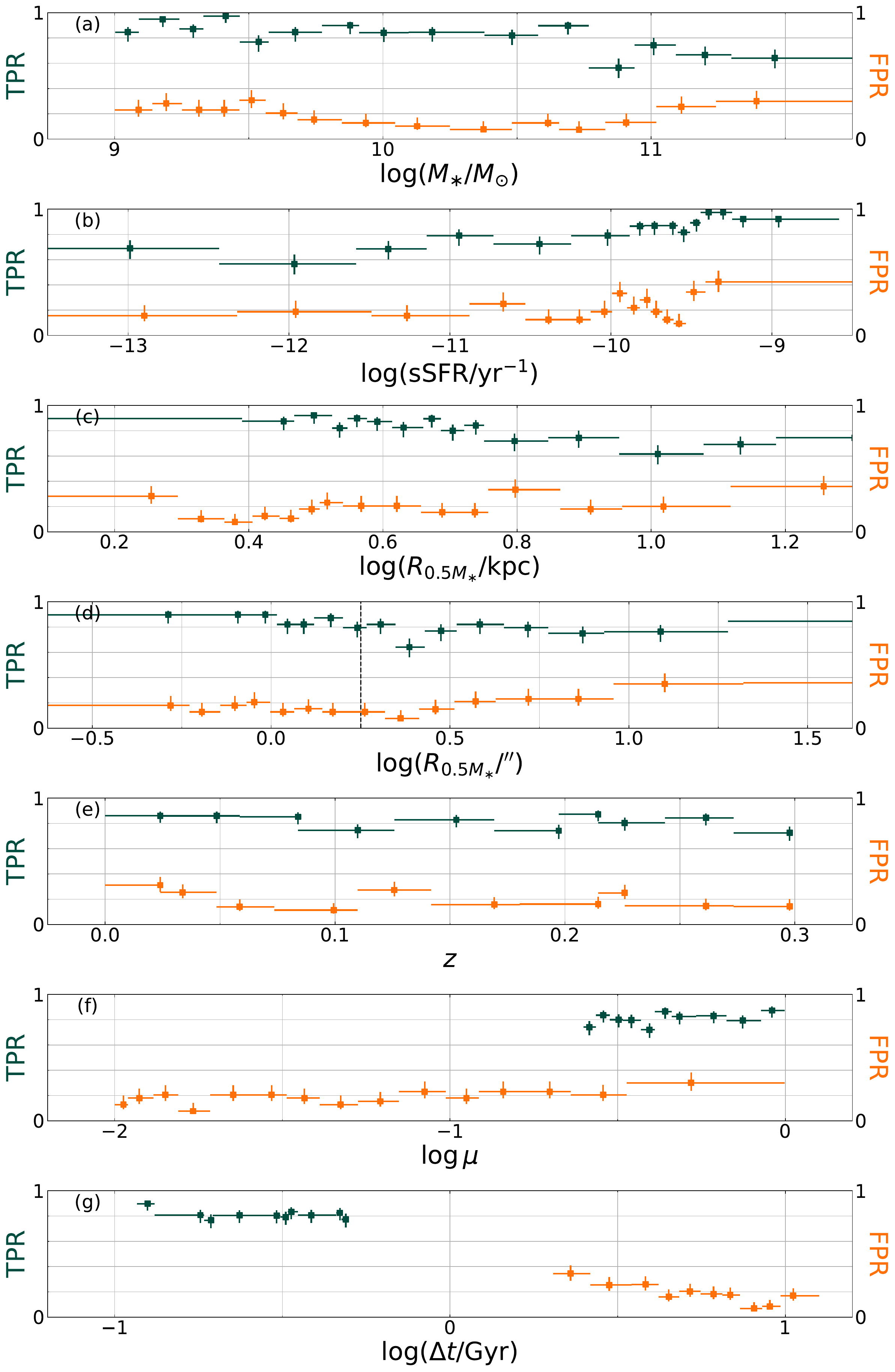}
  \caption{True positive rate (green) and false positive rate (orange) on the TNG training galaxies binned in (a) stellar mass, (b) specific star formation rate, (c) intrinsic half-mass radius, (d) angular half-mass radius, (e) merger stellar mass ratio, and (f) time since merger.}
  \label{fig:fpr-tpr}
\end{figure*}

To check for potential biases in the CNN classifier, Figures~\ref{fig:fmg-tng} and~\ref{fig:fpr-tpr} illustrate how CNN performance changes with galaxy properties. Figure~\ref{fig:fmg-tng} compares predicted with true merger fractions in the training galaxies binned in stellar mass, specific star formation rate, half mass radius (physical and angular), and redshift, while Figure~\ref{fig:fpr-tpr} shows true- and false-positive rates over the same properties in addition to merger mass ratio and time since merger. Table~\ref{table:tng-correlations} lists the Pearson correlation coefficients and their associated $p-$values for FPR, TPR, and, where applicable, $f_{\rm mg}$ excess ($\Delta f_{\rm mg}$; see Figure~\ref{fig:fmg-tng}). The correlations found among TPR suggest that the CNNs may be missing more true mergers at:
  \begin{enumerate}
  \item higher galaxy masses, which is unsurprising given the relative rarity of high-mass galaxies in the training set (see Figure~\ref{fig:tng-masses}). As the AGN and control samples are matched in stellar mass, this should not greatly affect the merger enhancements and depressions measured.
  \item lower specific star formation rates. This suggests that the decline of $f_{\rm mg}$ with sSFR may be steeper than shown in Figure~\ref{fig:f-mg-1d}b. This should not greatly affect our main result of a merger enhancement only in the blue cloud: if anything, additional mergers at lower sSFR could increase the magnitude of the differences seen.
  \item larger physical radii. This correlation is hardly surprising given that large-radius galaxies are rare, similar to high-mass galaxies, but it has a less obvious interpretation since we did not study $f_{\rm mg}$ evolution with physical size in the SDSS sample. As galaxy radius does correlate with stellar mass and overall SFR, we may expect merger completeness to decline moving towards larger $M_{\ast}$ and SFR in Figure~\ref{fig:enh-mstar-sfr}. Overall, this trend seems unlikely to affect our overall results.
  \end{enumerate}

\begin{table*}
  \centering
  \begin{tabular}{cccccccccc} \hline
    Property & & \multicolumn{2}{c}{TPR} & & \multicolumn{2}{c}{FPR} & & \multicolumn{2}{c}{$\Delta f_{\rm mg}$} \\ \hline 
             & & $r$ & $p$ & & $r$ & $p$ & & $r$ & $p$ \\ \hline\hline
    $\log(M_{\ast}/M_{\odot})$ & & ${\bf -0.73}$ & ${\bf 2.0 \times 10^{-3}}$ & & $-0.26$ & $0.36$ & & ${\bf -0.91}$ & ${\bf 2.3 \times 10^{-4}}$ \\ \hline
    $\log({\rm sSFR}/{\rm yr}^{-1})$ & & ${\bf 0.86}$ & ${\bf 3.8 \times 10^{-5}}$ & & $0.34$ & $0.22$ & & $0.66$ & $0.038$ \\ \hline
    $\log(R_{0.5M_{\ast}}/{\rm kpc})$ & & ${\bf -0.75}$ & ${\bf 1.3 \times 10^{-3}}$ & & $0.53$ & $0.042$ & & ${\bf -0.82}$ & ${\bf 3.8 \times 10^{-3}}$ \\ \hline
    $\log(R_{0.5M_{\ast}}/'')$ & & $0.07$ & $0.81$ & & $0.64$ & $0.01$ & & $0.20$ & $0.59$ \\ \hline
    $z$ & & $-0.44$ & $0.21$ & &  $-0.43$ & $0.22$ & & ${\bf -0.77}$ & ${\bf 8.8 \times 10^{-3}}$ \\ \hline
    $\log \mu$ & & $0.47$ & $0.17$ & & ${\bf 0.65}$ & ${\bf 9.1 \times 10^{-3}}$ & & -- & -- \\ \hline
    $\log(\Delta t/{\rm Gyr})$ & & $-0.45$ & $0.19$ & & ${\bf -0.86}$ & ${\bf 1.6 \times 10^{-3}}$ & & -- & -- \\ \hline
    
  \end{tabular}
  \caption{Pearson $r$ correlation coefficients and their associated $p-$values for true positive rate (TPR), false positive rate (FPR), and $f_{\rm mg}$ excess ($\Delta f_{\rm mg}$; see Figure~\ref{fig:fmg-tng}). Values with $p < 0.01$ are bolded.}
  \label{table:tng-correlations}
\end{table*}

  In terms of possible contamination, the only galaxy properties with which the merger predictions showed significant correlations with FPR were merger mass ratio and time since merger, neither of which can be measured in our SDSS sample. As discussed in depth in Section~\ref{subsec:results-validation}, these trends (coupled with the observed increase of $f_{\rm mg}$ with $B/T$, Figure~\ref{fig:f-mg-1d}c may indicate that the CNNs base their predictions more on overall morphology than on specific small-scale features.

  The residuals of predicted and true $f_{\rm mg}$ for the TNG testing set give a direct indication of where the CNNs may over- or underpredict merger fraction. Of the observable galaxy properties, the only property which significantly correlated with $\Delta f_{\rm mg}$ and had $\sigma_{\Delta f} \gtrsim \sigma_{f\ {\rm true}}$, i.e. changes in $\Delta f_{\rm mg}$ at least as large as the scatter in $f_{\rm mg, true}$, was stellar mass. Again, since the AGN hosts and controls are matched in mass, a slight underprediction of mergers at high stellar mass and slight overprediction at low stellar mass should not substantially affect our results.



\bsp	
\label{lastpage}
\end{document}